\documentclass[letterpaper,twocolumn,10pt]{article}
\usepackage{usenix2019_v3}

\usepackage{tikz}
\usepackage{mathtools}
\usepackage{amsfonts, amsthm} 
\usepackage{algorithm}
\usepackage[noend]{algpseudocode}
\usepackage[caption=false,font=scriptsize]{subfig}
\usepackage{filecontents}
\DeclareMathOperator*{\argmin}{arg\,min}

\usepackage{enumitem}
\usepackage{etoolbox}
\usepackage{tikz}
\usetikzlibrary{tikzmark}
\usetikzlibrary{calc}
\errorcontextlines\maxdimen

\newcommand{\ALGtikzmarkcolor}{black}% customise this, if you want
\newcommand{\ALGtikzmarkextraindent}{4pt}% customise this, if you want
\newcommand{\ALGtikzmarkverticaloffsetstart}{-.5ex}% customise this, if you want
\newcommand{\ALGtikzmarkverticaloffsetend}{-.5ex}% customise this, if you want
\makeatletter
\newcounter{ALG@tikzmark@tempcnta}

\newcommand\ALG@tikzmark@start{%
    \global\let\ALG@tikzmark@last\ALG@tikzmark@starttext%
    \expandafter\edef\csname ALG@tikzmark@\theALG@nested\endcsname{\theALG@tikzmark@tempcnta}%
    \tikzmark{ALG@tikzmark@start@\csname ALG@tikzmark@\theALG@nested\endcsname}%
    \addtocounter{ALG@tikzmark@tempcnta}{1}%
}

\def\ALG@tikzmark@starttext{start}
\newcommand\ALG@tikzmark@end{%
    \ifx\ALG@tikzmark@last\ALG@tikzmark@starttext
    \else
        \tikzmark{ALG@tikzmark@end@\csname ALG@tikzmark@\theALG@nested\endcsname}%
        \tikz[overlay,remember picture] \draw[\ALGtikzmarkcolor] let \p{S}=($(pic cs:ALG@tikzmark@start@\csname ALG@tikzmark@\theALG@nested\endcsname)+(\ALGtikzmarkextraindent,\ALGtikzmarkverticaloffsetstart)$), \p{E}=($(pic cs:ALG@tikzmark@end@\csname ALG@tikzmark@\theALG@nested\endcsname)+(\ALGtikzmarkextraindent,\ALGtikzmarkverticaloffsetend)$) in (\x{S},\y{S})--(\x{S},\y{E});%
    \fi
    \gdef\ALG@tikzmark@last{end}%
}

\apptocmd{\ALG@beginblock}{\ALG@tikzmark@start}{}{\errmessage{failed to patch}}
\pretocmd{\ALG@endblock}{\ALG@tikzmark@end}{}{\errmessage{failed to patch}}
\makeatother
\begin{document}
%-------------------------------------------------------------------------------

%don't want date printed
\date{}

\title{\Large \bf 2PS: High-Quality Edge Partitioning with Two-Phase Streaming}

%for single author (just remove % characters)
\author{
{\rm Ruben Mayer}\thanks{Both authors contributed equally to this paper.}\\
Technical University of Munich
\and
{\rm Kamil Orujzade}\textcolor{green!80!black}{\footnotemark[1]}\\
Technical University of Munich
\and
{\rm Hans-Arno Jacobsen}\\
Technical University of Munich
%{\rm Second Name}\\
%Second Institution
% copy the following lines to add more authors
% \and
% {\rm Name}\\
%Name Institution
} % end author

\maketitle

%-------------------------------------------------------------------------------
\begin{abstract}
%-------------------------------------------------------------------------------
Graph partitioning is an important preprocessing step to distributed graph processing. In edge partitioning, the edge set of a given graph is split into $k$ equally-sized partitions, such that the replication of vertices across partitions is minimized. Streaming is a viable approach to partition graphs that exceed the memory capacities of a single server. The graph is ingested as a stream of edges, and one edge at a time is immediately and irrevocably assigned to a partition based on a scoring function. However, streaming partitioning suffers from the \emph{uninformed assignment problem}: At the time of partitioning early edges in the stream, there is no information available about the rest of the edges. As a consequence, edge assignments are often driven by balancing considerations, and the achieved replication factor is comparably high. In this paper, we propose 2PS, a novel two-phase streaming algorithm for high-quality edge partitioning. In the first phase, vertices are separated into clusters by a lightweight streaming clustering algorithm. In the second phase, the graph is re-streamed and edge partitioning is performed while taking into account the clustering of the vertices from the first phase. Our evaluations show that 2PS can achieve a replication factor that is comparable to heavy-weight random access partitioners while inducing orders of magnitude lower memory overhead. 
\end{abstract}

%-------------------------------------------------------------------------------
\section{Introduction}
\label{sec:introduction}
In recent years, distributed graph processing frameworks, such as Pregel~\cite{pregel}, Giraph~\cite{giraph}, GPS~\cite{gps} and PowerGraph~\cite{powergraph}, have emerged as a way to perform complex graph analytics on massive real-world graphs on a cluster of servers. In a distributed graph processing framework, the graph analytics algorithms are provided in form of a vertex function (``think-like-a-vertex''), which is then executed iteratively until a convergence criterion is met~\cite{pregel}. In doing so, the different servers execute the vertex function on disjoint parts of the graph, which are called \emph{graph partitions}. While executing the vertex function, on each server, updates on vertex states are computed. If such updates refer to a vertex that is part of a different partition, communication between servers is initiated in order to synchronize the state, which is costly and increases the run-time of the graph processing task.
To minimize the required communication between servers that perform distributed graph processing, the goal of \emph{edge partitioning} is to divide the edges of the graph into $k$ equally-sized partitions such that the number of vertex replications is minimized. The graph partitioning problem is known to be NP-hard, and hence, can only be solved heuristically for large graphs~\cite{Zhang:2017:GEP:3097983.3098033}.

The current approaches to solve the edge partitioning problem can be categorized into two groups. \emph{Random-access} partitioners~\cite{Karypis:1998:FHQ:305219.305248, schlag2019scalable, Margo:2015:SDG:2824032.2824046, Zhang:2017:GEP:3097983.3098033, dne} load the complete graph into memory and then perform partitioning on it. In contrast to this, \emph{streaming} partitioners~\cite{Stanton:2012:SGP:2339530.2339722, dbh, Petroni:2015:HSP:2806416.2806424, 8416335} ingest the graph as a stream of edges, i.e., edge by edge, and assign each edge of the stream immediately and irrevocably to a partition. The streaming approach has the advantage that graphs can be partitioned with low memory overhead, as the edge set does not need to be kept in memory completely at any time. This way, the monetary costs of graph partitioning can be reduced, as smaller machines can be used for partitioning very large graphs. 

However, streaming edge partitioning yields a higher vertex replication than random-access approaches. We attribute this shortcoming to the \emph{uninformed assignment problem}~\cite{8416335}: A streaming partitioner only has a partial view of the graph structure. In particular, when processing an edge $e_i$ in the graph stream, the partitioner does not know the subsequent edges $e_{i+1}, e_{i+2}, ...$, but still has to decide to which partition it shall assign $e_i$. Current approaches to overcome the uninformed assignment problem either do not apply to the edge partitioning problem~\cite{Nishimura:2013:RGP:2487575.2487696} or require locality in the edge stream~\cite{8416335}.

In this paper, we provide a method for streaming edge partitioning that overcomes the uninformed assignment problem and yields a significantly lower replication factor than existing streaming solutions. Our contributions are threefold:

\begin{enumerate}[noitemsep]
	\item We propose 2PS, a novel two-phase streaming algorithm for edge partitioning. In the first phase, 2PS gathers information about the global graph structure by performing streaming clustering. In the second phase, clustering information is exploited to perform high-quality edge partitioning decisions. By combining two different algorithms, we exploit the flexibility of graph clustering and at the same time solve the more rigid edge partitioning problem.
	\item We provide a thorough theoretical analysis of 2PS regarding time and space complexity. Further, we prove that, on power-law graphs, the replication factor obtained with 2PS is strictly better than competing streaming partitioners.
	\item We perform extensive evaluations on large real-world graphs, showing the superior performance of 2PS compared to other streaming algorithms. On web graphs, 2PS yields a comparable replication factor to resource intensive random-access partitioners. 
\end{enumerate}

The rest of the paper is organized as follows. In Section~\ref{sec:background}, we formalize the problem of edge partitioning and introduce a model of stateful streaming partitioning. In Section~\ref{sec:approach}, we introduce our novel 2PS algorithm for high-quality stateful streaming edge partitioning. In Section~\ref{sec:analysis}, we perform a thorough theoretical analysis of 2PS regarding time and space complexity as well as replication factor on power-law graphs. In Section~\ref{sec:evaluations}, we perform extensive evaluations of 2PS on a variety of real-world graphs. Finally, in Section~\ref{sec:related} we discuss related work and then conclude the paper in Section~\ref{sec:conclusions}.
%-------------------------------------------------------------------------------
%-------------------------------------------------------------------------------
\section{Problem Analysis}
\label{sec:background}

\begin{figure}
\centering
  \includegraphics[width=0.65\linewidth]{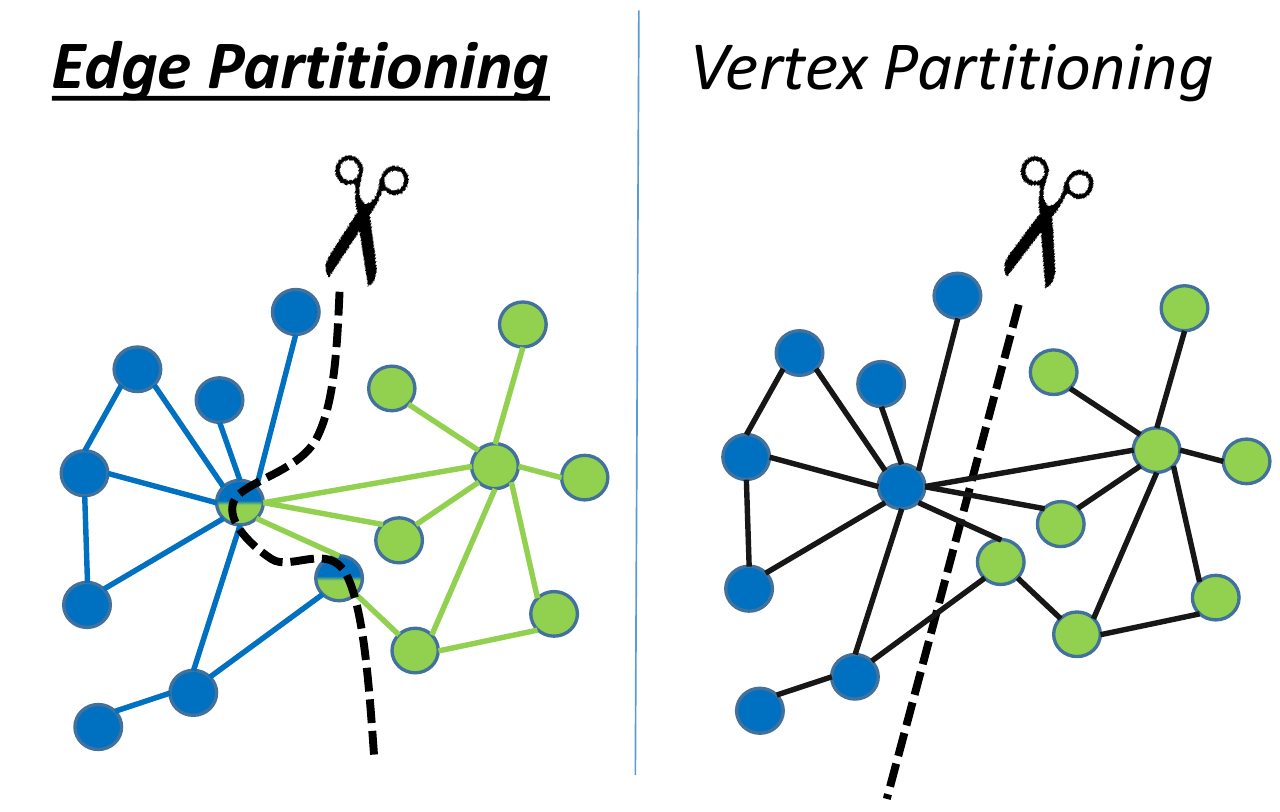}
%\vspace{-0.2cm}
  \caption{Edge partitioning vs. vertex partitioning.}
%\vspace{-0.3cm}
  \label{fig:cut_examples}
\end{figure}

\subsection{Edge Partitioning Problem}

\emph{Formalization.} The problem of \emph{edge partitioning} is commonly specified as follows (cf. also~\cite{Zhang:2017:GEP:3097983.3098033, Bourse:2014:BGE:2623330.2623660}). The $G = (V, E)$ is undirected or directed and consists of a set of vertices $V$ and a set of edges $E \subseteq V \times V$. Now, $E$ shall be split into  $k>1, k \in \mathbb{N}$ partitions $P = \{p_1, ..., p_k\}$  such that $\bigcup_{i=1,...,k} p_i = E$ and  $p_i \cap p_j = \emptyset, i \neq j$, while a balancing constraint is met: $\forall p_i \in P : |p_i| \leq \alpha  * \frac{|E|}{k} $ for a given $\alpha \geq 1, \alpha \in \mathbb{R}$. The balancing constraint ensures that the largest partition does not exceed the expected number of edges multiplied by an imbalance factor $\alpha$ that limits the acceptable imbalance. We define $V(p_i)=\{x \in V | \exists y \in V : (x,y) \in p_i \lor (y,x) \in p_i \}$ as the set of vertices covered by a partition $p_i \in P$, i.e., the set of vertices that are incident to an edge in $p_i$. The optimization objective of edge partitioning is to minimize the \emph{replication factor} RF($p_1, \dots, p_k$)$\ = \frac{1}{|V|} \sum_{i=1,...,k}{|V(p_i)|}$.\footnote{``Communication volume''~\cite{Margo:2015:SDG:2824032.2824046} is an alternative metric that can directly be computed from the replication factor by subtracting 1.}

\emph{Interpretation.} The replication of a vertex on multiple partitions induces synchronization overhead in distributed graph processing. In particular, vertex state needs to be synchronized between distributed compute nodes that hold different partitions. The lower the replication factor, the lower is the synchronization overhead in distributed graph processing. Numerous studies~\cite{Zhang:2017:GEP:3097983.3098033, dne} prove that there is a direct correlation between replication factor in edge partitioning and run-time of distributed graph processing. 

Figure~\ref{fig:cut_examples} depicts an example of a graph that is divided into $k=2$ partitions with edge partitioning (left side) vs. vertex partitioning (right side). Contrary to edge partitioning, in vertex partitioning, the vertex set of the graph is divided into partitions and edges are cut~\cite{Stanton:2012:SGP:2339530.2339722}. For graphs with a power-law distributed vertex degree, edge partitioning is considered more communication-efficient than vertex partitioning (cf. Bourse et al.~\cite{Bourse:2014:BGE:2623330.2623660}). This is because power-law graphs have a small number of vertices with a very high degree; cutting through these vertices via edge partitioning is very effective in separating the graph with a low replication factor.

\begin{figure}
\centering
  \includegraphics[width=0.94\linewidth]{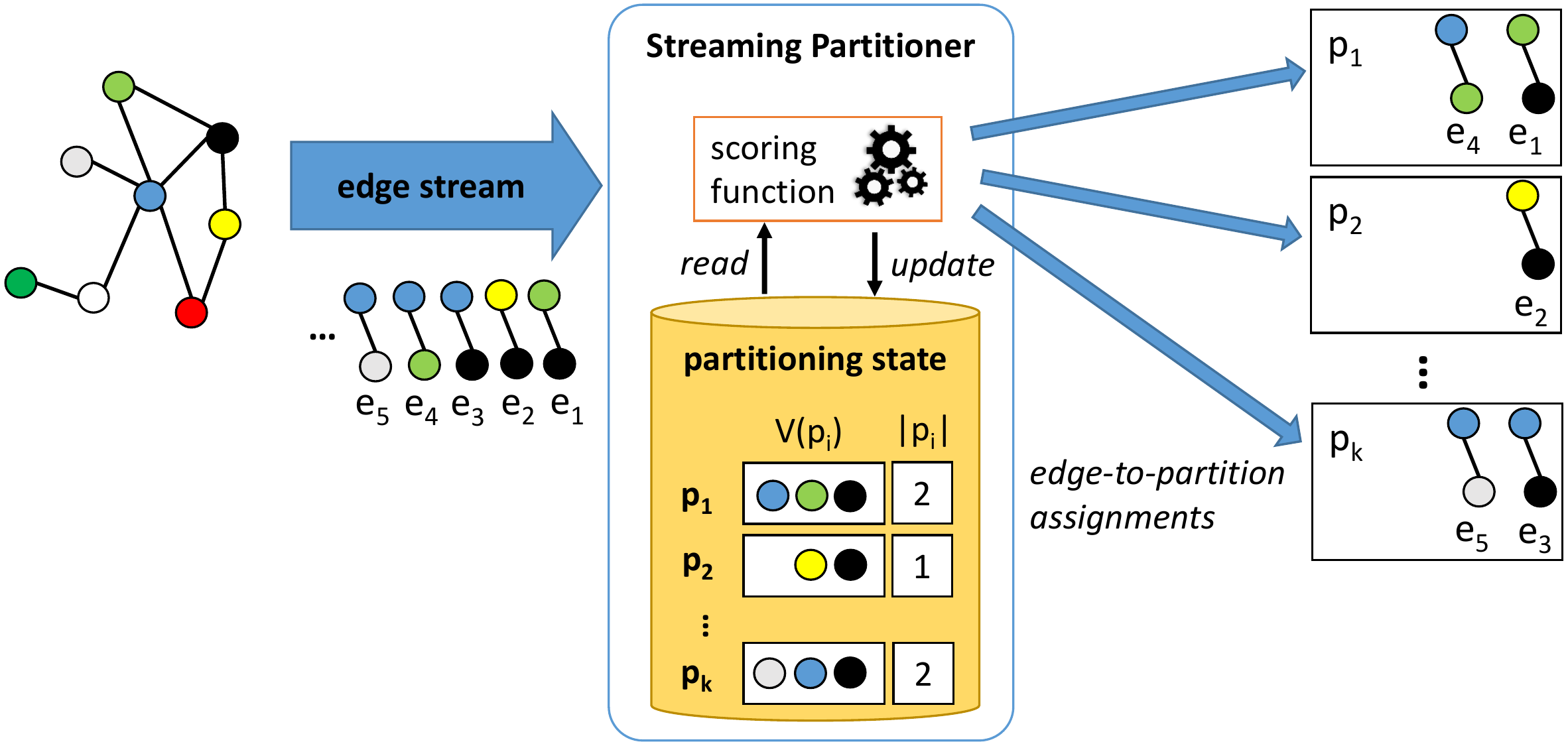}
%\vspace{-0.2cm}
  \caption{A model of stateful streaming edge partitioning.}
%\vspace{-0.3cm}
  \label{fig:stateful_streaming}
\end{figure}

\subsection{Streaming Edge Partitioning}

\emph{Streaming edge partitioning} promises to reduce the memory overhead of edge partitioning. In particular, space complexity of streaming partitioning is independent of the number of edges in the graph. To do so, in streaming edge partitioning, the graph is ingested edge by edge (one edge at a time), and each edge is immediately assigned to a partition (cf. Figure~\ref{fig:stateful_streaming}). The decision to assign an edge to a partition is performed based on a \emph{scoring function} that can take into account \emph{graph properties} (e.g., the known or estimated degrees of the incident vertices of the edge~\cite{dbh, Petroni:2015:HSP:2806416.2806424}) as well as \emph{partitioning state} (e.g., the vertex cover sets of the partitions and the current size of the partitions~\cite{Petroni:2015:HSP:2806416.2806424}). For stateful streaming partitioning, the size of the partitioning state is limited to $\mathcal{O}(|V|)$, i.e., only vertex-related information, to avoid that the memory overhead grows with the number of edges in the graph.

As a practical example, HDRF~\cite{Petroni:2015:HSP:2806416.2806424} is a streaming partitioner that assigns an edge $e =(u,v)$ to the partition $p$ which maximizes a scoring function $C^{\mathit{HDRF}}(u,v,p) = C_{\mathit{REP}}(u,v,p) + C_{\mathit{BAL}}(p)$, where $C_{\mathit{REP}}(u,v,p)$ is a degree-weighted replication score and $C_{\mathit{BAL}}(p)$ is a balancing score. $C_{\mathit{REP}}(u,v,p)$ is highest if both vertices $u$ and $v$ incident to an edge $e$ are in the vertex cover set of the same partition~$p$; $C_{\mathit{BAL}}(p)$  is highest when $p$ is the smallest partition (i.e., contains least number of edges).

\subsection{Shortcomings of Streaming Partitioning} 

In stateful streaming partitioning, partitioning state contains the information about which vertex is already replicated on which partition. When partitioning the $i$-th edge in the stream, the partitioning state considers the assignments of all previous edges $e_h, h < i$ in the graph stream to partitions. However, it does not encompass information about \emph{future} edges that have not been visited in the graph stream yet. Hence, \emph{``early''} edges in the stream are assigned to partitions with only little partitioning state available, which may lead to poor quality of such assignments, i.e., a high replication factor. We introduced this problem in Section~\ref{sec:introduction} as the \emph{uninformed assignment problem}. 

Furthermore, streaming partitioning may be sensitive to the ordering of the graph stream. In HDRF~\cite{Petroni:2015:HSP:2806416.2806424}, the authors observe that balancing can be negatively influenced if the stream order exposes a high locality of vertices. They advise to \emph{shuffle} the graph before ingestion. While shuffling mitigates adverse stream ordering, it induces the problem of poor memory access locality on the vertex-based partitioning state which may slow down the execution (cf. also~\cite{Wei:2016:SGP:2882903.2915220}).

%-------------------------------------------------------------------------------
%-------------------------------------------------------------------------------
\section{Two-Phase Streaming Edge Partitioning}
\label{sec:approach}

\begin{figure}
\centering
  \includegraphics[width=\linewidth]{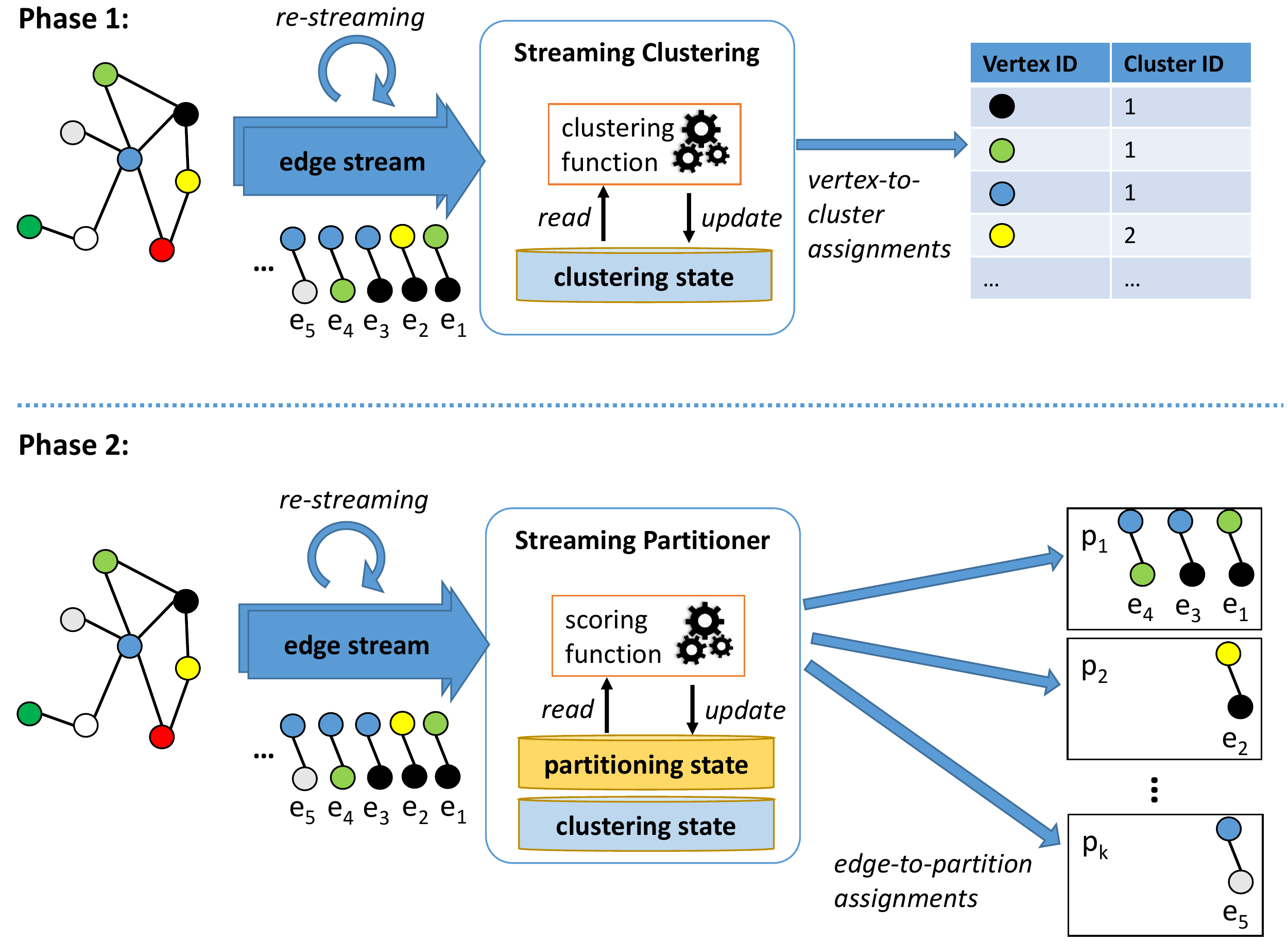}
%\vspace{-0.2cm}
  \caption{2PS: Approach Overview.}
%\vspace{-0.3cm}
  \label{fig:approach_overview}
\end{figure}

In our work, we tackle the uninformed assignment problem while staying faithful to the streaming paradigm. We achieve this by introducing a novel two-phase streaming algorithm, called 2PS (cf. Figure~\ref{fig:approach_overview}). %As a side effect of performing multiple passes through the edge stream, the sensitivity of 2PS to the stream ordering is only low, making it unnecessary to shuffle the edge stream before ingestion.

The two phases of 2PS consist of (1) a streaming clustering phase, in which vertices are assigned to clusters based on their neighborhood relationships, and (2) a streaming edge partitioning phase, where clustering information is exploited to achieve a low replication factor.

\subsection{Phase 1: Clustering}

The aim of the clustering phase is to gather information about the graph structure in order to guide the subsequent streaming edge partitioning phase. We observe that a group of vertices should be replicated on the same partition if there are many edges between vertices of that group, i.e., the group is densely connected. This way, many edges can be assigned to a partition while only few vertices are added to the vertex cover set of that partition, leading to a low overall replication factor.

Finding sets of vertices that are densely connected to each other, while only having few connections to other vertices that are outside of that group, is a graph problem that is well known as \emph{clustering} or \emph{community detection}~\cite{Newman8577, FORTUNATO201075}. Formally, the goal of clustering is to maximize \emph{modularity}: $Q = \frac{1}{2*|E|}\sum_{u \in V}\sum_{v \in V}(w_{uv} - \frac{d_u*d_v}{2*|E|})*\delta(u,v)$, where $w_{uv}$ denotes the number of edges between vertices $u$ and $v$, $d_u$ the degree (number of incident edges) of a vertex $u$, and $\delta(u, v) = 1$ if $u$ and $v$ belong to the same cluster, and 0 otherwise~\cite{hollocou2017streaming}.\footnote{An intuitive explanation of modularity is provided by Newman~\cite{Newman8577}: \emph{``The modularity is, up to a multiplicative constant, the number of edges falling within groups minus the expected number in an equivalent network with edges placed at random.''}} 

Despite of some similarities, clustering and edge partitioning have a different nature and, hence, are addressed with different algorithms~\cite{Newman8577}. In particular, \emph{clustering is a less constrained problem} than partitioning. First, the size of the different clusters does not have to be balanced, i.e., clusters are allowed to have different sizes (although they may have to adhere to a maximum size). Contrary to this, in edge partitioning, every partition has to cover an equal (up to the imbalance factor $\alpha$) number of edges. 
The second difference between clustering and edge partitioning is that the number of clusters is not predetermined, but originates from the structure of the graph. Contrary to this, in edge partitioning, the number of partitions is determined by the user. The less constrained nature of clustering allows for divising a more efficient and flexible streaming algorithm. 

Another advantage of clustering over edge partitioning is the possibility to change the assignment of a vertex to a cluster multiple times in one single pass through the edge stream. A vertex of degree $d$ is incident to $d$ edges, and therefore, is visited $d$ times in one single pass through the edge stream. Every time a vertex is visited, its assignment to a cluster can be refined, taking into account new information that has been gathered since the last time the vertex was visited. Contrary to this, in edge partitioning, in a single pass through the edge list, every edge is only visited once and is immediately assigned to a partition. It is not trivially possible to revoke an edge-to-partition assignment at a later point in time when more information about the graph structure is accumulated. To re-assign edges to different partitions would require to track the mapping of \emph{edges} to partitions. Such mapping, however, can not be kept in memory for graphs with a large edge set.

\subsubsection{Streaming Clustering Algorithm}

\begin{algorithm}[t]
\caption{2PS Phase 1: Clustering}
\begin{algorithmic}[1]
%\footnotesize
\State int[] \emph{d} \Comment{vertex degrees}
\State int[] \emph{vol} \Comment{cluster volumes}
\State int[] \emph{v2c} \Comment{map of vertex\_id to cluster\_id}
\State int \emph{max\_vol} \Comment{maximum cluster volume}
\State int  \emph{next\_id} $\gets$ 0 \Comment{id of next new cluster}
\vspace{0.2cm}
\Procedure{streamingClustering}{}
	\State \emph{max\_vol} $\gets$ $\frac{2*|E|}{k} * 0.5$
	\State \texttt{performStreamingPass}() \Comment{first pass}
	\State \emph{max\_vol} $\gets$ \emph{max\_vol} $ * 2$
	\State \texttt{performStreamingPass}() \Comment{second pass}
\EndProcedure

\vspace{0.2cm}
\Procedure{performStreamingPass}{}
\For{\textbf{each} $e \in $ edge\_stream}
	\For {\textbf{each} $v \in e$}
		\If{\emph{v2c}[v] = NULL} %\Comment{create a new cluster}
			\State \emph{v2c}[$v$] $\gets$ next\_id
			\State \emph{vol}[\emph{next\_id}] $\gets$ \emph{vol}[\emph{next\_id}] + \emph{d}[$v$]
			\State \emph{next\_id} $\gets$ \emph{next\_id} + 1
		\EndIf
	\EndFor
	%\State $c_1 \gets $\emph{v2c}[$e$.first]; $c_2 \gets $\emph{v2c}[$e$.second]
	\If{\emph{vol}[\emph{v2c}[$v$]] $\leq$ \emph{max\_vol} $\forall v \in e$}
	\State $v_{\mathit{s}} \gets$ $v_i \in e : $\emph{vol}[\emph{v2c}[$v_i$]] $\leq$ \emph{vol}[\emph{v2c}[$v_j$]], $v_j\in e$ 
	\State $v_{\mathit{l}} \gets$ $v_j \in e : v_j \neq v_{\mathit{s}}$ 
		\If{\emph{vol}[\emph{v2c}[$v_l$]] $+$ \emph{d}[$v_s$] $\leq$ \emph{max\_vol}}
			\State \emph{v2c}[$v_s$] $\gets$ \emph{v2c}[$v_l$] 
			\State \emph{vol}[\emph{v2c}[$v_l$]] $\gets$ \emph{vol}[\emph{v2c}[$v_l$]] $+$ \emph{d}[$v_s$]]
			\State \emph{vol}[\emph{v2c}[$v_s$]] $\gets$ \emph{vol}[\emph{v2c}[$v_s$]] $-$ \emph{d}[$v_s$]]
		\EndIf
	\EndIf
\EndFor
\EndProcedure
%\State 
\end{algorithmic}
\label{alg:clustering}
\end{algorithm}

In clustering, every vertex is assigned to a cluster id, with the number of clusters not being predetermined. The aim of clustering in 2PS is to later pre-partition edges that are incident to two vertices of the same cluster, i.e., to find \emph{intra-cluster edges}. For this reason, in our clustering algorithm, we limit the cluster size such that the number of intra-cluster edges does not exceed the maximum size of a partition. To further improve clustering quality, we apply re-streaming~\cite{Nishimura:2013:RGP:2487575.2487696}, i.e., we perform two subsequent passes through the edge list; in each pass, we execute the same streaming clustering algorithm, allowing clustering decisions from the first stream pass to be revised in the second pass. Between the first and the second pass, we relax the cluster volume constraint to achieve both balanced cluster sizes and high clustering quality\footnote{We observed in our experiments that performing more than two stream passes does not significantly improve the quality of clustering.}.

Our streaming clustering algorithm (Algorithm~\ref{alg:clustering}) is an extension of an algorithm proposed by Hollocou et al.~\cite{hollocou2017streaming}. The algorithm consumes the edge stream, processing it edge by edge (line 12). It first checks for each incident vertex of the current edge $e$ if it already is assigned to a cluster. If this is not the case, a new cluster is created and the vertex is assigned to it (lines 13--17). In other words, clusters are constructed in a bottom-up manner, starting with small clusters that only contain a single vertex. Now, the algorithm compares the cluster volumes of the incident vertices of $e$. The cluster volume is defined as the sum of degrees of all vertices of a cluster. The incident vertex of $e$ that is currently assigned to the cluster with the lower volume is moved to the cluster of the incident vertex of $e$ with the higher volume. In other words, when processing an edge, an incident vertex migrates from a smaller cluster to a larger cluster (lines 18--24). However, such migration is only allowed if the new volume of the larger cluster does not exceed a volume bound (line 21).

Compared to Hollocou's original algorithm~\cite{hollocou2017streaming}, our extensions are the following. First of all, Hollocou's algorithm requires to ingest a shuffled input edge streaming order to avoid that all vertices are assigned to the same cluster. Second, despite of ingesting a shuffled edge list, Hollocou's algorithm does not guarantee that a maximum number of intra-cluster edges per cluster is kept. In 2PS, we compute the degree of each vertex upfront (if not already known) and use the actual vertex degree instead of the partial degree in order to compute cluster volumes. This way, the adverse effect of sorted edge streams is mitigated; high-degree vertices get a higher volume from the start, avoiding that all vertices are ``sucked'' into one giant cluster. Furthermore, we enforce an explicit volume cap on the clusters. As we consider the actual degree of vertices instead of the partial degree, we can enforce such volume cap effectively. 

%\newpage

\begin{algorithm}
\caption{2PS Phase 2: Streaming Partitioning}
\begin{algorithmic}[1]
%\footnotesize
\State int[] \emph{d} \Comment{vertex degrees (from Phase 1)}
\State int[] \emph{vol} \Comment{cluster volumes (from Phase 1)}
\State int[] \emph{v2c} \Comment{map of vertex\_id to cluster\_id (from Phase 1)}
\State int[] \emph{c2p} \Comment{map of cluster\_id to partition\_id}
\State int[] \emph{vol\_p} \Comment{sum of volumes of clusters per partition} 
\State int[][] \emph{v2p} \Comment{vertex\_id to partition\_id replication bit matrix}
\vspace{0.1cm}
\Procedure{streamingPartitioning}{}
	\State \texttt{mapClustersToPartitions}()
	\State \texttt{prepartitionEdges}() 
	\State \texttt{partitionRemainingEdges}()
\EndProcedure
\vspace{0.1cm}

\Procedure{mapClustersToPartitions}{}
\State sort clusters by volume (descending)
\For{\textbf{each} cluster $c$} (from largest to smallest)
	\State \emph{target\_p} $\gets$ $\argmin_{p_i \in P}$\emph{vol\_p}[$p_i$]	%\Comment{smallest aggregate volume}
	\State \emph{c2p}[$c$] $\gets$ \emph{target\_p}
\EndFor
\EndProcedure
\vspace{0.1cm}

\Procedure{prepartitionEdges}{}
\For{\textbf{each} $e \in $ edge\_stream}
	\State \emph{c\_1} $\gets$ \emph{v2c}[$e$.first]
	\State \emph{c\_2} $\gets$ \emph{v2c}[$e$.sec]
		\If{\emph{c\_1} = \emph{c\_2} \textbf{OR} \emph{c2p}[\emph{c\_1}] = \emph{c2p}[\emph{c\_2}]} %\Comment{both vertices are in same cluster}
		\State \emph{target\_p} $\gets$  \emph{c2p}[\emph{c\_1}]
		\If{|\emph{target\_p}| $\geq \alpha  * \frac{|E|}{k} $}
			\For{\textbf{each} $p_i \in P : |p_i| < \alpha  * \frac{|E|}{k} $} %\Comment{compute HDRF score}
				\State \emph{score} $\gets$ $\mathit{scoring\_function}$($e$, $p_i$) 
				\If{\emph{score} $>$ \emph{bestScore}}
					\State \emph{bestScore} $\gets$ \emph{score}
					\State \emph{target\_p} $\gets p_i$ 
				\EndIf
			\EndFor
		\EndIf
		\State \emph{v2p}[$e$.first][\emph{target\_p}] $\gets$ true
		\State \emph{v2p}[$e$.sec][\emph{target\_p}] $\gets$ true
		\State \texttt{output}: $e$ assigned to \emph{target\_p}
	\EndIf
\EndFor
\EndProcedure
\vspace{0.1cm}

\Procedure{partitionRemainingEdges}{}
\For{\textbf{each} $e \in $ edge\_stream}
	\State \emph{c\_1} $\gets$ \emph{v2c}[$e$.first]
	\State \emph{c\_2} $\gets$ \emph{v2c}[$e$.sec]
		\If{\emph{c\_1} = \emph{c\_2} \textbf{OR} \emph{c2p}[\emph{c\_1}] = \emph{c2p}[\emph{c\_2}]} 
		\State \textbf{continue} \Comment{skip pre-partitioned edge}
	\EndIf
	\State \emph{bestScore} $\gets 0$
	\State \emph{target\_p} $\gets$ NULL
	\For{\textbf{each} $p_i \in P : |p_i| < \alpha  * \frac{|E|}{k} $} %\Comment{compute HDRF score}
		\State \emph{score} $\gets$ $\mathit{scoring\_function}$($e$, $p_i$) 
		\If{\emph{score} $>$ \emph{bestScore}}
			\State \emph{bestScore} $\gets$ \emph{score}
			\State \emph{target\_p} $\gets p_i$ 
		\EndIf
	\EndFor
	\State \emph{v2p}[$e$.first][\emph{target\_p}] $\gets$ true
	\State \emph{v2p}[$e$.sec][\emph{target\_p}] $\gets$ true
	\State \texttt{output}: $e$ is assigned to \emph{target\_p}
\EndFor
\EndProcedure
%\vspace{0.1cm}
%
%\Procedure{HDRF\_Score}{$u$, $v$, $p$}
%	%\State \emph{sum\_of\_degrees} $\gets d[u] + d[v]$
%	\State \emph{normalized\_degree\_u} $\gets \frac{d[u]}{d[u] + d[v]}$
%	\State \emph{normalized\_degree\_v} $\gets \frac{d[v]}{d[u] + d[v]}$
%\EndProcedure
\end{algorithmic}
\label{alg:partitioning}
\end{algorithm}

\subsection{Phase 2: Partitioning}

The edge partitioning algorithm (Algorithm~\ref{alg:partitioning}) consists of three subsequent steps. First, clusters are mapped to partitions. Second, a subset of edges are pre-partitioned by exploiting the clustering of their incident vertices. Third, remaining edges are partitioned by stateful streaming edge partitioning. In the following, we describe each step in detail.

\textbf{\emph{Step 1: Mapping Clusters to Partitions.}} Our objective in the first step is to map clusters to partitions, such that the total volume of clusters across partitions is balanced. We model this problem as an instance of the classical \texttt{Makespan Scheduling Problem on Identical Machines} (\texttt{MSP-IM}). The problem can be defined as follows~\cite{graham1969bounds}:

\begin{quote}Given a set of $k$ machines $M_1, ..., M_k$ and a list of $n$ jobs $j_1, ..., j_n$ with corresponding run-time $a_1,  ..., a_n$, assign each job to a machine such that the makespan is minimized\footnote{The makespan is the time taken until all jobs have been completed.}.\end{quote}

We apply our cluster assignment problem to \texttt{MSP-IM} as follows. Partitions are corresponding to ``machines'', clusters are corresponding to ``jobs'' and volumes of the clusters are corresponding to ``run-times'' of the jobs. The optimization goal is to minimize the cumulative volume of the largest partition.

\texttt{MSP-IM} is NP-hard~\cite{Ullman:1975:NSP:1739944.1740138}, so that we solve it by approximation. The \emph{sorted list scheduling algorithm} by Graham~\cite{graham1969bounds} is a $\frac{4}{3}$-approximation of \texttt{MSP-IM}, i.e., its result is at most $\frac{4}{3}$ times as large as the true optimum. Sorted list scheduling works as follows: First, sort the jobs by decreasing runtime, then assign job by job from the sorted list to the least loaded machine. Applied to our cluster assignment problem, this means that the clusters are sorted by decreasing volume (Algorithm~\ref{alg:partitioning}, line 12) and then assigned one by one to the currently least loaded partition (lines 13 to 15). Now, it is guaranteed that the most loaded partition is at most $\frac{4}{3}$ as large as it would be in the true optimal assignment.

\textbf{Step 2: Pre-Partition Edges by Exploiting Vertex Clusters.}
In the second step, we exploit the clustering of vertices to partition a subset of edges. Our goal is to perform pre-partitioning \emph{\underline{perfectly}}, i.e., after pre-partitioning, \emph{every vertex incident to a pre-partitioned edge shall, if possible, only be covered by \textbf{one} single partition}. To do so, the pre-partitioning algorithm performs one pass through the complete edge stream (Algorithm~\ref{alg:partitioning}, line 17). For each edge $e=(e.$first$, e.$second$)$, it checks if both incident vertices $e$.first and $e$.second are either in the same cluster or their clusters are assigned to the same partition $p$ (cf. Step 1 discussed above). In this case, $e$ is applicable to pre-partitioning and shall be assigned to $p$ (lines 18 to 21). If $p$ is already occupied to its maximum capacity $\alpha * \frac{|E|}{k}$, $e$ is assigned to a different partition instead, using stateful streaming edge partitioning.

\textbf{Step 3: Partition Remaining Edges.}
Partitioning the remaining edges is done via stateful streaming edge partitioning. Remaining edges are all edges not assigned to partitions in the pre-partitioning step. These are edges between vertices of different clusters that are mapped to different partitions. After pre-partitioning, the partitioning state is already filled with information on vertex covers of the different partitions for the pre-partitioned edges. This state is exploited in the partitioning of the remaining edges to improve their placement. We enforce a hard balancing cap, i.e., we \emph{guarantee} that no partition gets more than $\alpha * \frac{|E|}{k}$ edges assigned\footnote{Remarkably, such strict balancing guarantee is not provided or implemented in many of the existing edge partitioners, e.g., NE/SNE~\cite{Zhang:2017:GEP:3097983.3098033}, DNE~\cite{dne} and HDRF~\cite{Petroni:2015:HSP:2806416.2806424}.}. In our implementation of 2PS, we adopt the scoring function from HDRF~\cite{Petroni:2015:HSP:2806416.2806424} to perform stateful streaming partitioning. %However, 2PS does not depend on a specific scoring function; other scoring functions~\cite{Stanton:2012:SGP:2339530.2339722, 8416335} could also be used.

For partitioning the remaining edges, the algorithm performs a pass through the edge stream (Algorithm~\ref{alg:partitioning}, line 32), processing it edge by edge. First, it determines whether the edge has already been pre-partitioned by checking the conditions for pre-partitioning (incident vertices are in the same cluster or in clusters that are mapped to the same partition). If the conditions for pre-partitioning are met, the edge is skipped (lines 33 to 36). Else, stateful streaming partitioning is applied (lines 37 to 46). In doing so, partitions that have already reached their maximum capacity of $\alpha * \frac{|E|}{k}$ edges are ignored (line 39). 

After Step 3 is finished, all edges have been assigned to partitions and none of the partitions has more than $\alpha * \frac{|E|}{k}$ edges. This concludes the 2PS algorithm.

\subsection{Discussion of Design Choices}

Here, we discuss our design choices and compare them to possible alternatives. 

First, we could have attempted to devise a simpler re-streaming approach for edge partitioning, where the same partitioning function is used in multiple streaming passes through the edge set. In each streaming pass, edge assignments from the previous pass could be revised while exploiting the partitioning state. This idea has been proposed and analyzed for the vertex partitioning problem by Nishimura and Ugander~\cite{Nishimura:2013:RGP:2487575.2487696}. However, for edge partitioning, re-streaming would be more complex. When reassigning an edge $e = (u,v)$, i.e., changing its partition from a previous stream pass, we would need to remember the ``old'' partition $p\_{old}$ of $e$ to be able to update the partitioning state, which involves testing if $u$ or $v$ should be removed from the vertex replication set of $p\_{old}$. However, remembering all edge-to-partition assignments implies a space complexity of $\mathcal{O}(|E|)$. Our solution of applying different techniques in different streaming phases does not suffer from this problem; our space complexity is independent of the number of edges (see Section~\ref{sec:space complexity}).

Second, instead of using the actual degree of vertices, we could have relied on a partial vertex degree in the clustering phase, which avoids the need for an additional pass through the stream in case the vertex degrees are not known a priori. However, relying on partial degrees only works if the edge stream is shuffled, as Hollocou et al.~\cite{hollocou2017streaming} report. The problem of shuffled edge lists is that access to vertex state would have low locality, leading to many cache misses, which induces a higher run-time. Indeed, we found that it is faster to make an additional stream pass to compute the real vertex degrees instead of using a shuffled edge list.

%-------------------------------------------------------------------------------
%-------------------------------------------------------------------------------
\section{Theoretical Analysis}
\label{sec:analysis}
\subsection{Time Complexity}

We analyze each phase of 2PS separately. Phase 1, specified in Algorithm~\ref{alg:clustering}, performs two passes through the edge set. In each pass, a constant number of operations is performed on each edge. Hence, the time complexity of the first phase is in $\mathcal{O}(|E|)$. Phase 2, specified in Algorithm~\ref{alg:partitioning}, consists of three steps. First, clusters are mapped to partitions in decreasing volume order. To sort clusters by volume is in $\mathcal{O}(|V| * \log |V|)$, as in the worst case, there are as many clusters as vertices (note that, in natural graphs, we can expect the number of clusters to be orders of magnitude smaller than the number of vertices). Each cluster is assigned to the currently least loaded partition, which can be performed in $\mathcal{O}(|V| * \log k)$ time, provided that we keep the $k$ partitions sorted by their accumulated volume while assigning clusters to them. Second, edges are pre-partitioned, such that edges whose incident vertices both are in clusters of the same partition are assigned to that partition. This is a constant-time operation per edge, resulting in $\mathcal{O}(|E|)$ time complexity. Third, the remaining edges are partitioned using HDRF scoring, which is done in $\mathcal{O}(|E|*k)$ time. In summary, the second phase of 2PS has a time complexity of $\mathcal{O}(|E|*k)$, as $|E| >> |V|$. Summing up, the total time complexity of 2PS is in $\mathcal{O}(|E|*k)$, i.e., linear in the number of edges and in the number of partitions.

\subsection{Space Complexity}
\label{sec:space complexity}

We analyze the data structures used in 2PS. In Algorithm~\ref{alg:clustering}, we use arrays to store the vertex degrees, cluster volumes and a mapping of vertices to clusters. Each of these data structures has a space complexity of $\mathcal{O}(|V|)$. In Algorithm~\ref{alg:partitioning}, besides these arrays, we use additional arrays to map the clusters to partitions and to keep the volumes of clusters per partition. These arrays all have a space complexity of $\mathcal{O}(|V|)$. Finally, we use a vertex-to-partition replication matrix, which has a space complexity of $\mathcal{O}(|V|) * k)$. Hence, the overall 2PS algorithm has a space complexity of  $\mathcal{O}(|V|) * k)$. In particular, the space complexity is independent of the number of edges in the graph.

\subsection{Replication Factor}
\label{sec:rf}

We prove that the average replication factor of 2PS is strictly lower than the average replication factor of HDRF according to Petroni et al.~\cite{Petroni:2015:HSP:2806416.2806424}. HDRF is currently the partitioning algorithm with the lowest proven bound on average replication factor \emph{among streaming partitioners}. 

We construct our proof as follows. Let the set of edges $|E|$ be separated into two disjoint subsets $E_{pp}$ and $E_{HDRF}$, such that $E = E_{pp} \cup E_{HDRF}$ and $E_{pp} \cap E_{HDRF} = \emptyset$. $E_{pp}$ is the set of edges assigned to partitions via pre-partitioning based on vertex clustering and $E_{HDRF}$ is the set of edges assigned to partitions based on HDRF scoring. Now, we extend the notation of vertex cover sets as follows. Let $V_{pp}  = \{x \in V | \exists y \in V : (x,y) \in E_{pp} \lor (y,x) \in E_{pp} \}$ the set of vertices covered by $E_{pp}$, and analogously,  $V_{HDRF}  = \{x \in V | \exists y \in V : (x,y) \in E_{HDRF} \lor (y,x) \in E_{HDRF} \}$ the set of vertices covered by $E_{HDRF}$. We analyze the average replication factor of three different, disjoint vertex sets. Let $V_A$ be the set of vertices that are in $V_{pp}$, but not in $V_{HDRF}$: $V_A = V_{pp} \setminus V_{HDRF}$. Further, let $V_B$ be the set of vertices that are in $V_{HDRF}$, but not in $V_{pp}$: $V_B = V_{HDRF} \setminus V_{pp}$. Finally, let $V_C$ be the set of vertices that are both in $V_{pp}$ and in $V_{HDRF}$: $V_C = V_{pp} \cap V_{HDRF}$. Notice that $V_A$, $V_B$, and $V_C$ are pairwise disjoint. Now, we prove that for each set $V_A$, $V_B$, and $V_C$, the average replication factor achieved with 2PS is equal or better than the average replication factor achieved with HDRF.

$V_A$: Pre-partitoning only replicates every vertex in $V_{A}$ one single time to one single target partition, which is the partition the corresponding cluster of the vertex is mapped to. These vertices have a perfect replication factor of 1, which is lower than the replication factor that is guaranteed by HDRF partitioning.

$V_B$: Vertices in $V_B$ are only touched by HDRF partitioning in the last streaming phase of 2PS. Hence, their average replication factor will be the same.

$V_C$: Vertices in $V_C$ are both touched by pre-partitioning as well as by HDRF partitioning (first and second pass, respectively, of the 2PS partitioning phase). Along the lines of the proof provided by Petroni et al.~\cite{Petroni:2015:HSP:2806416.2806424}, we differentiate whether a vertex $v \in V_C$ is a high-degree vertex (hub vertex) or a low-degree (non-hub) vertex. If $v$ is a hub vertex, it will be replicated by HDRF partitioning to every partition. The replication of $v$ by 2PS pre-partitioning, hence, does not have any effect on the replication factor. If $v$ is a non-hub vertex and HDRF processes an edge between $v$ and a hub vertex $u_h$, HDRF would only replicate $v$ if it is the first time it observes $v$. Now, because $v \in V_{pp}$, there must already be partitioning state of $v$ gathered from the pre-partitioning pass, i.e., $v$ cannot be observed for the first time, and hence, will not be replicated by HDRF. If $v$ is a non-hub vertex and HDRF processes an edge between $v$ and another non-hub vertex $u_l$, HDRF may replicate $v$. As pre-partitioning has also replicated $v$, the replication factor of HDRF could not have been better than pre-partitioning in 2PS. 

Summing up, the average replication factor of vertices in $V_A$ is strictly lower than the bounds provided by HDRF, while the average replication factor of vertices in $V_B$ and $V_C$ is equal to the bounds of HDRF. Hence, 2PS achieves a lower bound on average replication factor than HDRF, given that there are vertices in $V_A$---in particular, vertices that have only intra-cluster edges and no inter-cluster edges.
\hfill $\qed$

%-------------------------------------------------------------------------------%-------------------------------------------------------------------------------
\section{Evaluations}
\label{sec:evaluations}

\subsection{Setup}

\textbf{Data Sets.} We perform a set of performance evaluations on real-world graphs (cf. Table~\ref{tab:graphs}) of different scales and origin. TW\footnote{https://snap.stanford.edu/data/twitter-2010.html}~\cite{Kwak:2010:TSN:1772690.1772751, snapnets} and FR\footnote{https://snap.stanford.edu/data/com-Friendster.html}~\cite{6413740, snapnets} are social networks; IT\footnote{http://law.di.unimi.it/webdata/it-2004/}~\cite{BMSB, BRSLLP, BoVWFI}, UK\footnote{http://law.di.unimi.it/webdata/uk-2007-05/}~\cite{BMSB, BRSLLP, BoVWFI}, GSH\footnote{http://law.di.unimi.it/webdata/gsh-2015/}~\cite{BMSB, BRSLLP, BoVWFI} and WDC\footnote{http://webdatacommons.org/hyperlinkgraph/} are web graphs. 

\renewcommand{\arraystretch}{1.15}
\begin{table}
{\scriptsize
	\begin{center}
		\begin{tabular}{l|l|l|l|l}
			\hline
			Name & \textbf{$|V|$} & \textbf{$|E|$} & Size & Type \\	
			\hline
			it-2004 (IT) & 41 M & 1.2 B & 9 GiB & Web \\
			twitter-2010 (TW) & 42 M & 1.5 B & 11 GiB & Social \\
			com-friendster (FR) & 66 M & 1.8 B & 14 GiB & Social \\
			uk-2007-05 (UK) & 106 M & 3.7 B & 28 GiB & Web \\
			gsh-2015 (GSH) & 988 M & 33 B & 248 GiB & Web \\
			wdc-2014 (WDC) & 1.7 B & 64 B & 478 GiB & Web \\	
			\hline
		\end{tabular}
	\end{center}
}
%\vspace{-5pt}
\caption{Real-world graph datasets. Size refers to the graph representation as binary edge list with 32-bit vertex ids.}
\label{tab:graphs}
%\vspace{-5pt}
\end{table}

\textbf{Baselines.} From the group of streaming partitioners, we compare to HDRF~\cite{Petroni:2015:HSP:2806416.2806424}, DBH~\cite{dbh} and SNE (a streaming version of the random-access partitioning algorithm NE~\cite{Zhang:2017:GEP:3097983.3098033}). We chose these baselines for the following reasons. HDRF delivers low replication factor at modest run-time and memory overhead. DBH is based on hashing and is the fastest streaming partitioner with the lowest memory overhead. Finally, SNE delivers a lower replication factor than HDRF, albeit at a significantly higher run-time and memory overhead. Other streaming partitioners (Greedy~\cite{powergraph}) are outperformed by our chosen baseline partitioners, or they are too slow to partition large graphs in a reasonable time (ADWISE~\cite{8416335}).

While random-access partitioners generally induce a large memory overhead, they are a challenging baseline in terms of replication factor and run-time. From the group of random-access partitioners, we compare to NE~\cite{Zhang:2017:GEP:3097983.3098033} and DNE~\cite{dne}. These are currently the partitioners that achieve the lowest replication factor (NE) and the best scalability (DNE). Other random-access partitioners take too much memory or run-time to partition the real-world graphs on our evaluation platform (KaHiP~\cite{schlag2019scalable}) or they are outperformed by DNE in terms of replication factor and run-time (Spinner~\cite{spinner}, ParMetis~\cite{Karypis:1998:FHQ:305219.305248}, XtraPulp~\cite{xtrapulp}, Sheep~\cite{Margo:2015:SDG:2824032.2824046}). 

For a fair comparison, we re-implemented the HDRF algorithm in C++ and use the same implementation of HDRF also for edge partitioning in the second phase of 2PS. We also re-implemented DBH, using the same framework that we developed for HDRF and 2PS. For NE, SNE and DNE, we use the reference implementation of the respective authors. All implementations that we use, except for DNE, ingest the graph in the same binary input format (i.e., binary edge list with 32-bit vertex ids). DNE ingests an ASCII edge list---there is no option for binary input provided in the reference implementation of DNE.

\textbf{Evaluation Platform.} We perform all experiments on a server with 4 x 16 Intel(R) Xeon(R) CPU E5-4650 0@2.70GHz, 503 GiB of main memory and 1 x 4.4 TB HDD disk. The operating system is Ubuntu 18.04.2 LTS. 

\textbf{Experimental Settings.} We used the following settings throughout all experiments. The balancing constraint $\alpha$ was set to 1.05. Further, we configured the system parameters of the baseline partitioner according to the authors' recommendations as follows. We configured HDRF with a setting of $\lambda = 1.1$. For SNE, we used a cache size of $2*|V|$. For DNE, we set an expansion ratio of 0.1.

For each experiment, we re-compiled the partitioners with optimal settings, as far as compilation flags were offered. For 2PS, HDRF and DNE, we set the maximum number of partitions to the numbers used in the corresponding experiment. Further, we compiled DNE with 32-bit vertex ids. This way, the memory overhead of the partitioners was minimized as far as possible.

Finally, for DNE, we adapted the number of threads per process as follows. In DNE, for each partition, a separate process is spawned. As our evaluation machine offers 64 hardware threads, we tried to reach a total number of 64 threads, if possible. This means that each process gets $\lceil\frac{64}{k}\rceil$ threads.

\subsection{Performance on Real-World Graphs}
\label{sec:realworld}

We perform our experiments for $k = \{4, 32, 128, 256\}$ partitions. We repeat each experiment 3 times and report the mean value along with error bars that show the standard deviation. For each set of experiments, we perform an initial warm-up run which does not count into the results. To use our limited server time efficiently, we aborted experiments (reporting an \emph{out of time} error) when a partitioner used more than 6 times the run-time of 2PS. 

The key performance metrics we report are replication factor, partitioning run-time, and memory overhead. We also track balancing. In most cases, the balancing constraint is met by all partitioners; if this is not the case, we report the measured imbalance factor $\alpha$ in the plot.

\begin{figure*}
	\centering	
	%\subfloat[CAPTION]{BILDERCODE}\qquad
	\subfloat[IT: Replication factor.]{\includegraphics[width=0.295\textwidth]{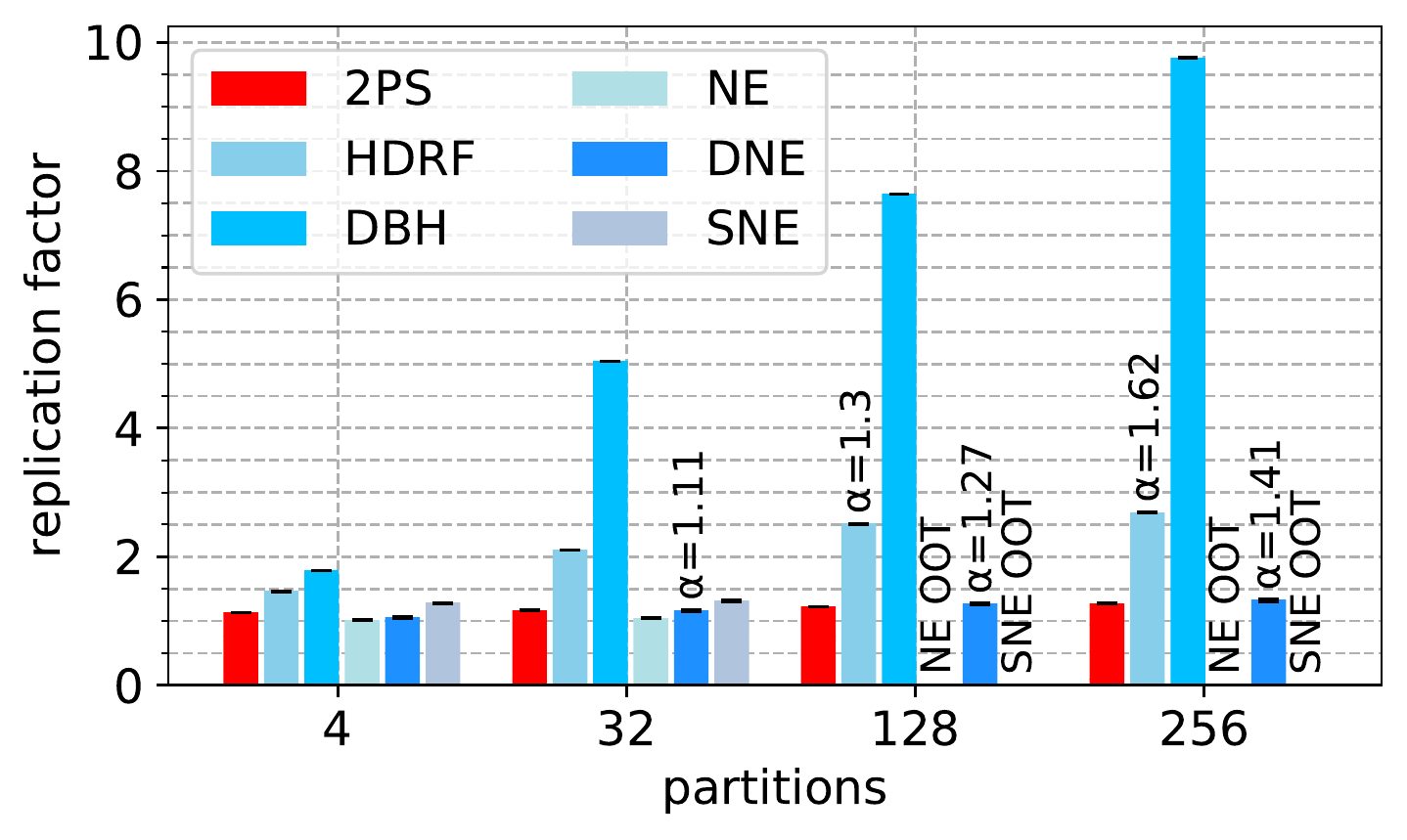}}
	\subfloat[IT: Run-time.]{\label{b}   \includegraphics[width=0.295\textwidth]{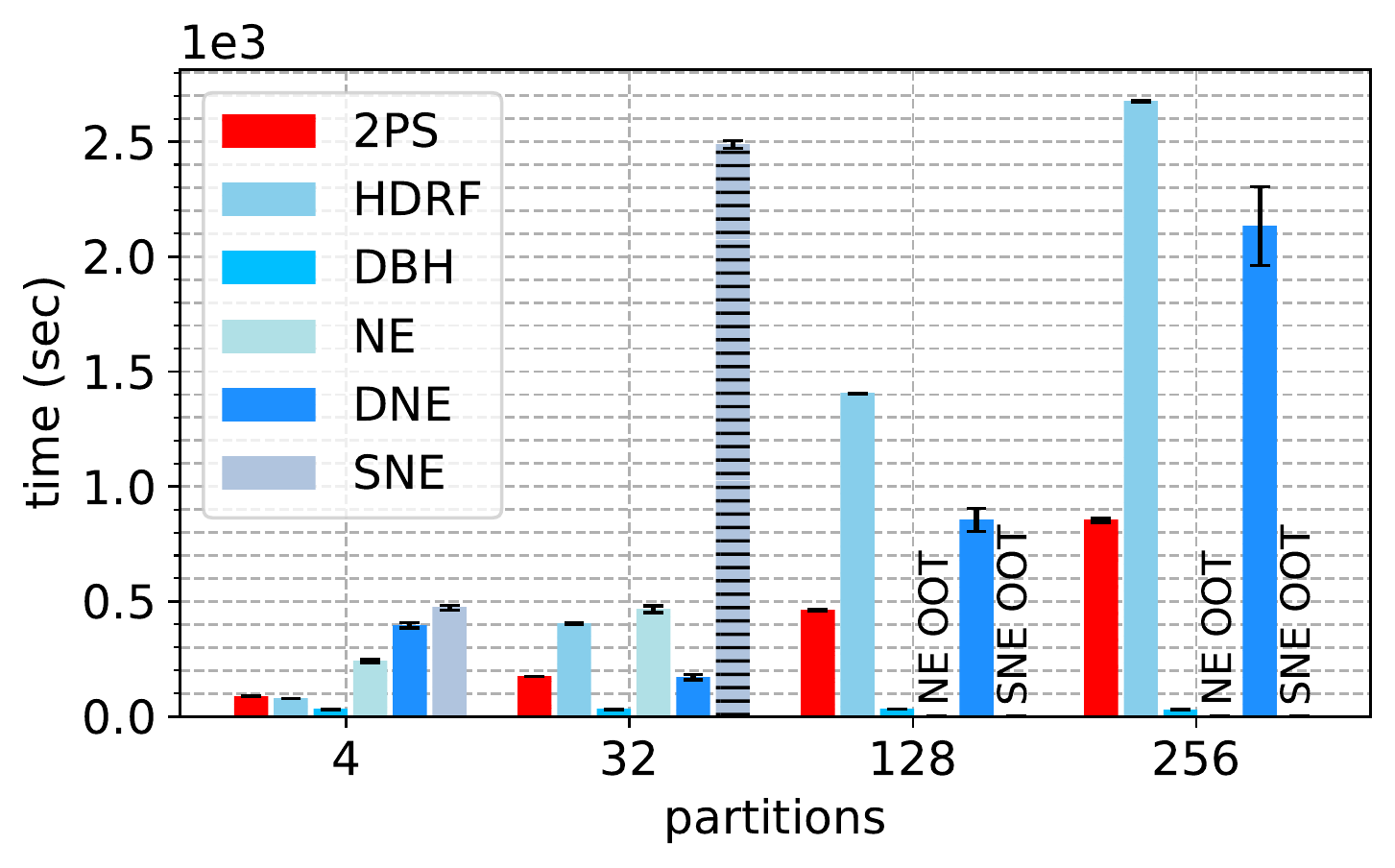}} 
	\subfloat[IT: Memory overhead.]{\label{c}   \includegraphics[width=0.295\textwidth]{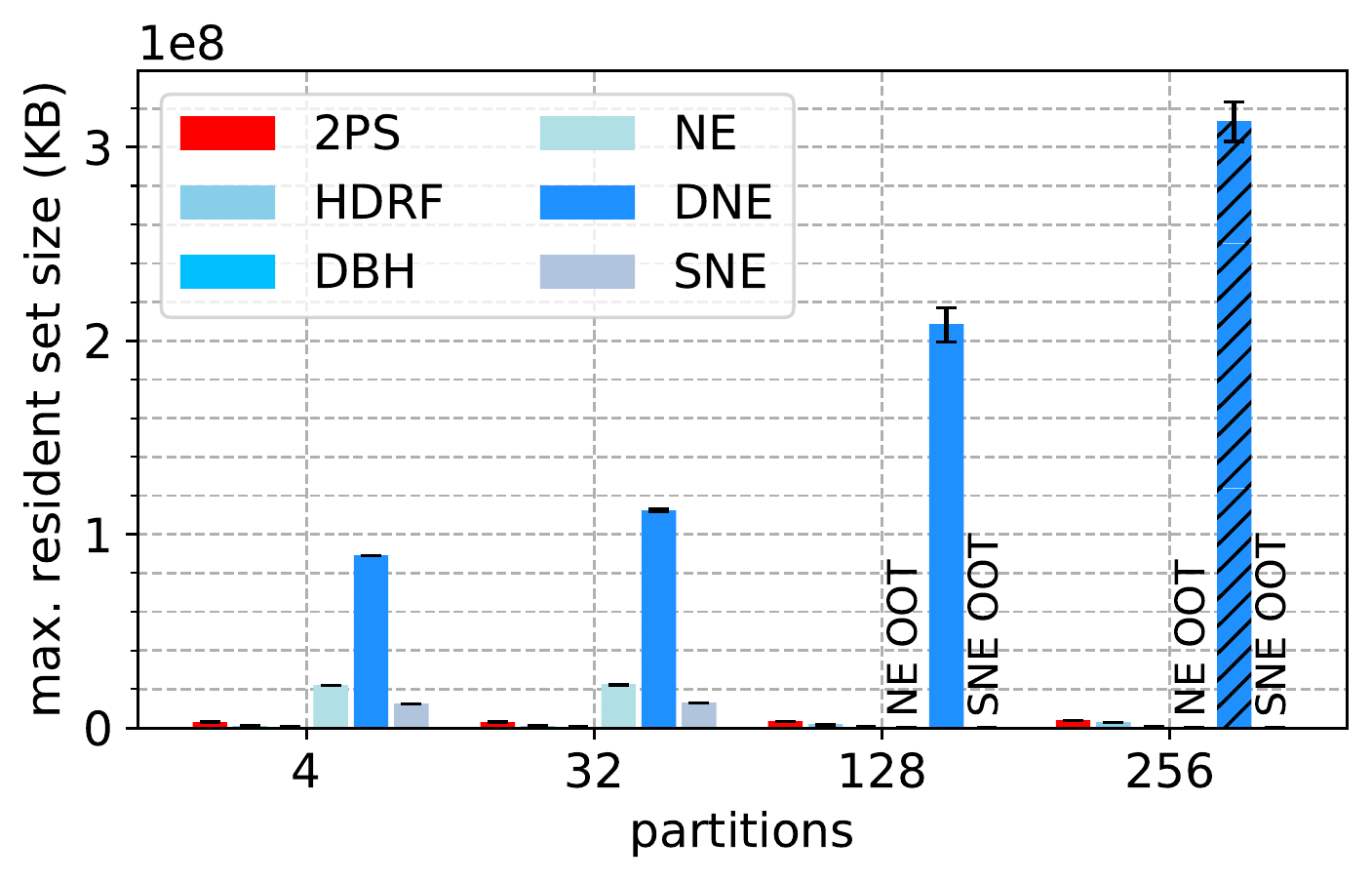}}\\
	\vspace{-0.45cm}
	\subfloat[TW: Replication factor.]{\label{a}   \includegraphics[width=0.295\textwidth]{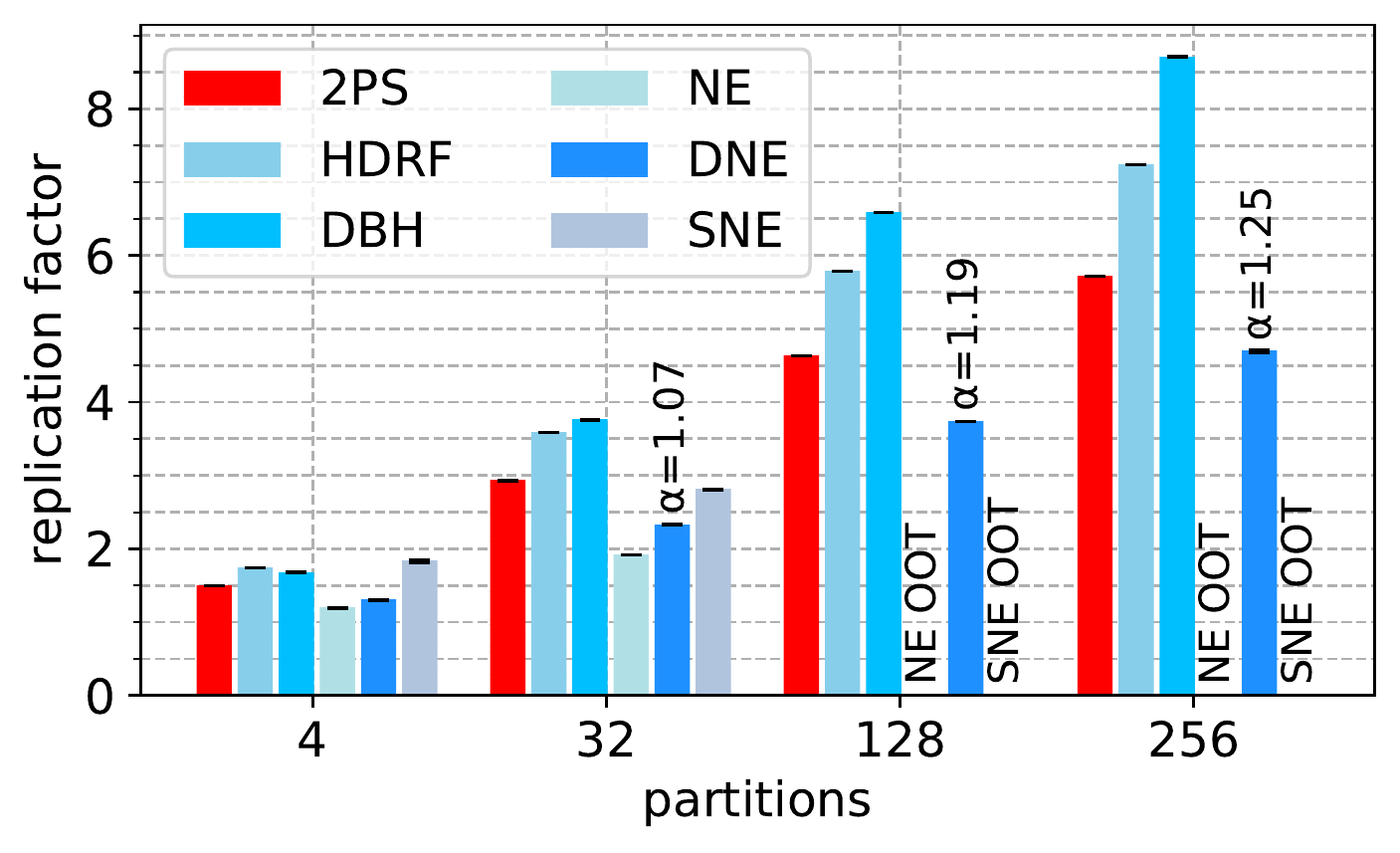}}
	\subfloat[TW: Run-time.]{\label{b}   \includegraphics[width=0.295\textwidth]{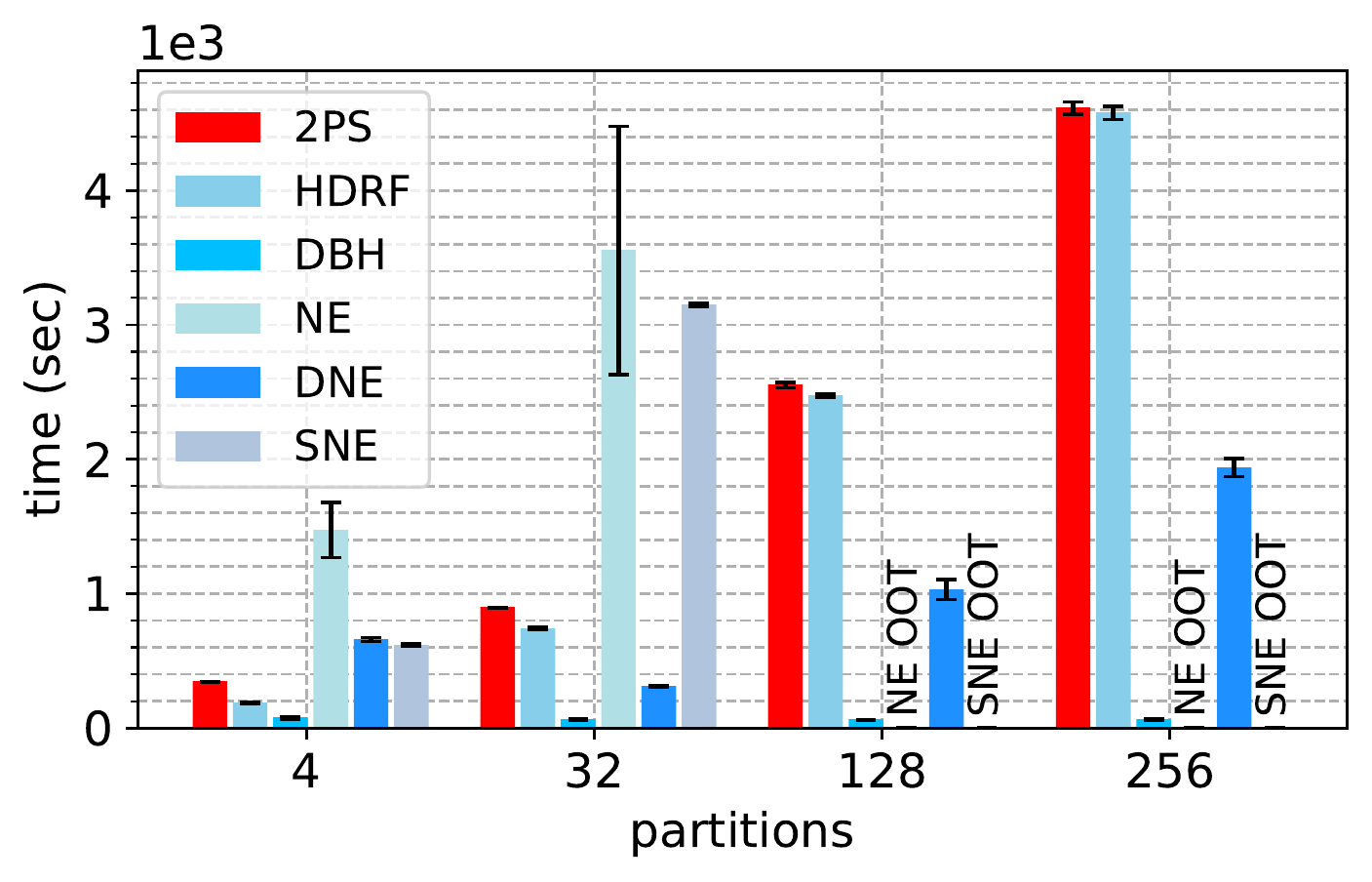}}
	\subfloat[TW: Memory overhead.]{\label{c}   \includegraphics[width=0.295\textwidth]{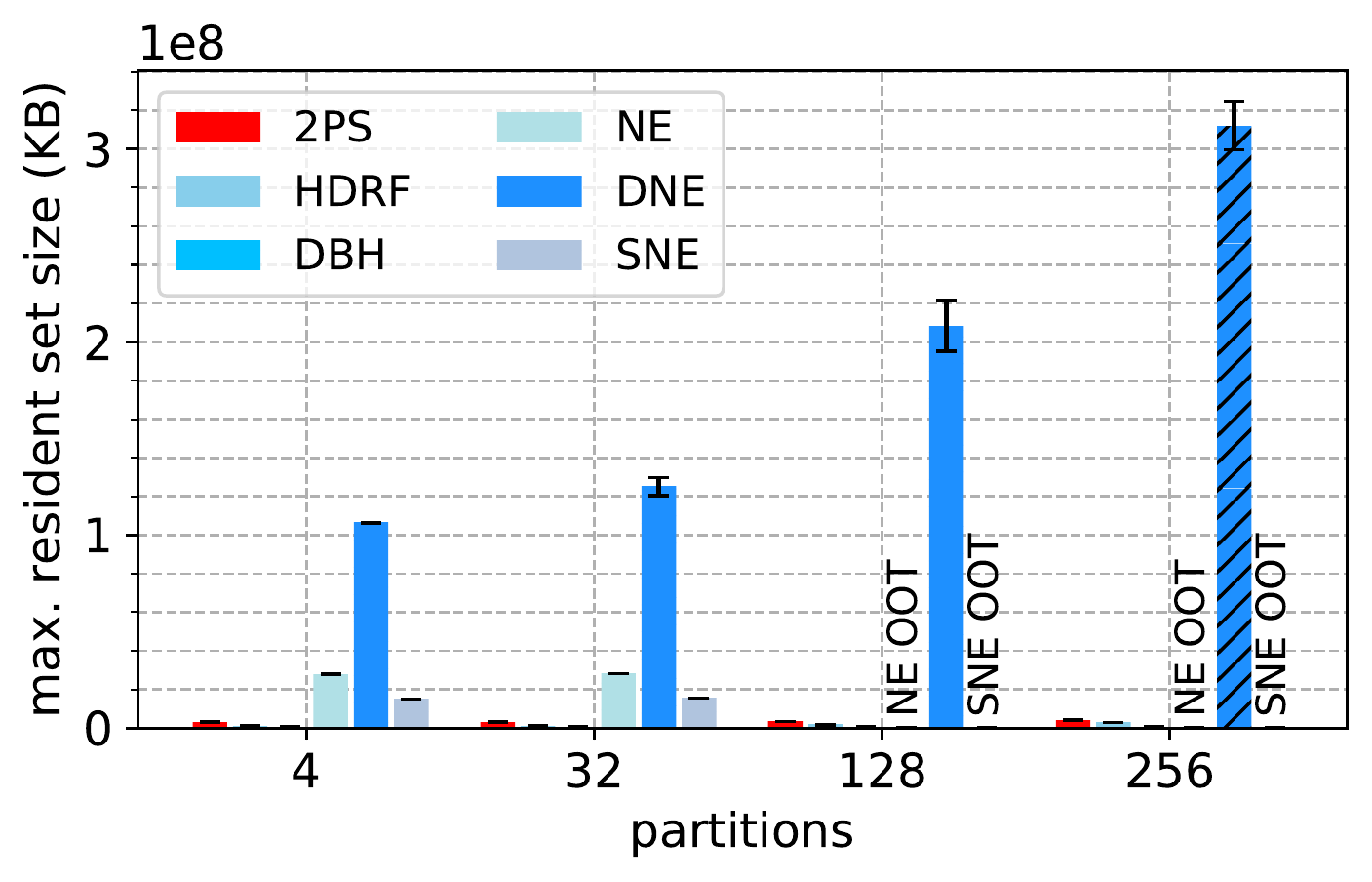}} \\
	\vspace{-0.45cm}
	\subfloat[FR: Replication factor.]{\label{a}   \includegraphics[width=0.295\textwidth]{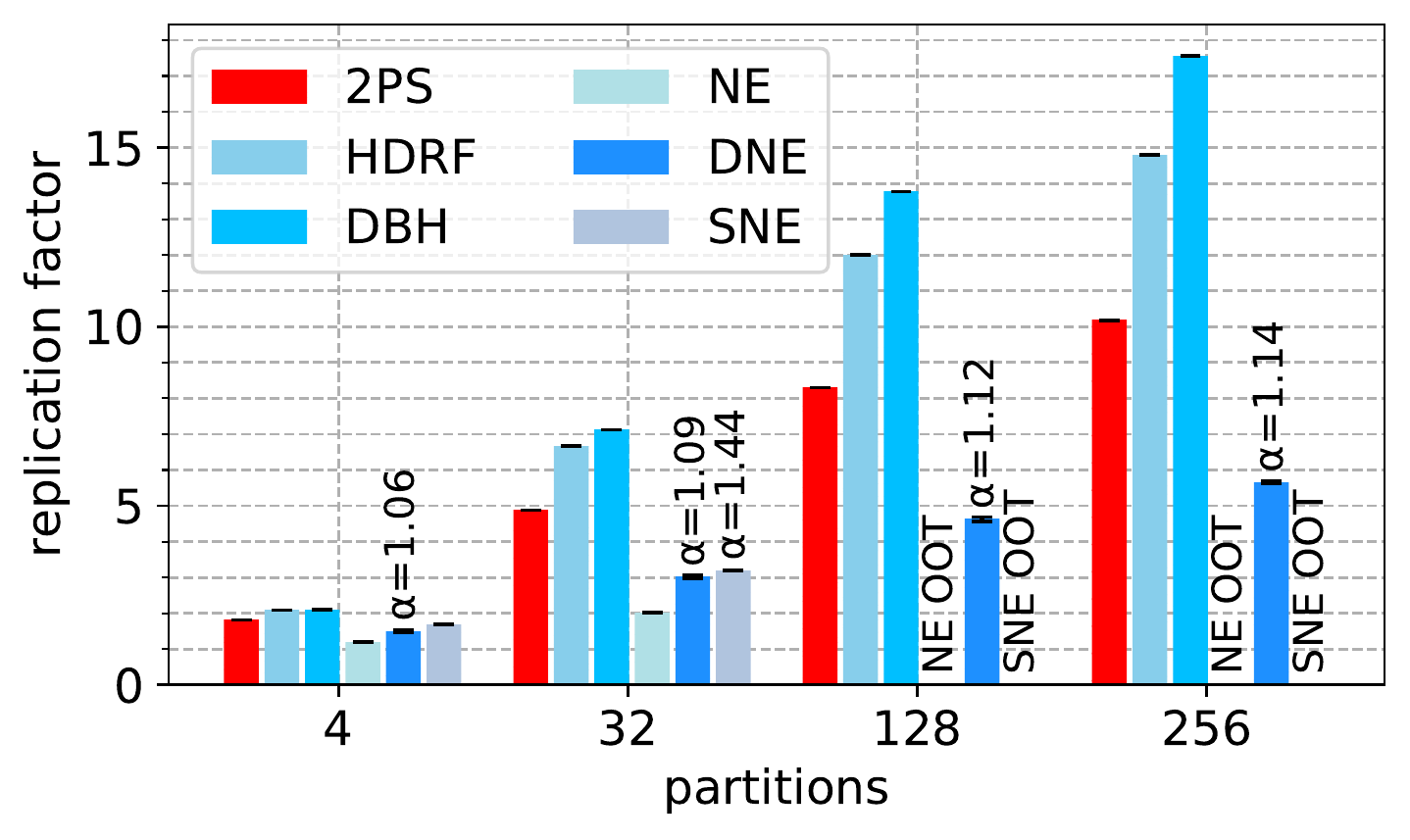}}
	\subfloat[FR: Run-time.]{\label{b}   \includegraphics[width=0.295\textwidth]{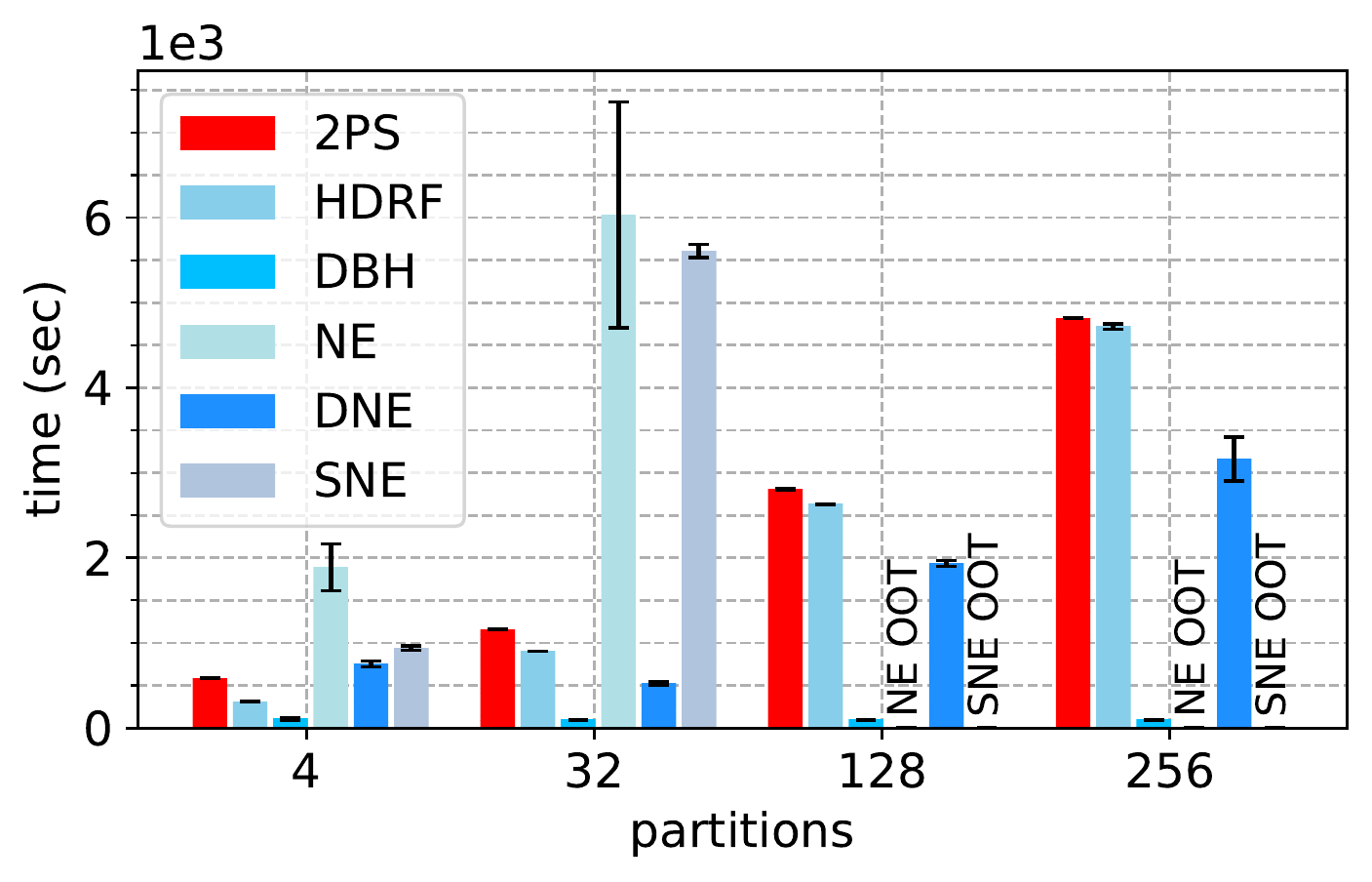}}
	\subfloat[FR: Memory overhead.]{\label{c}   \includegraphics[width=0.295\textwidth]{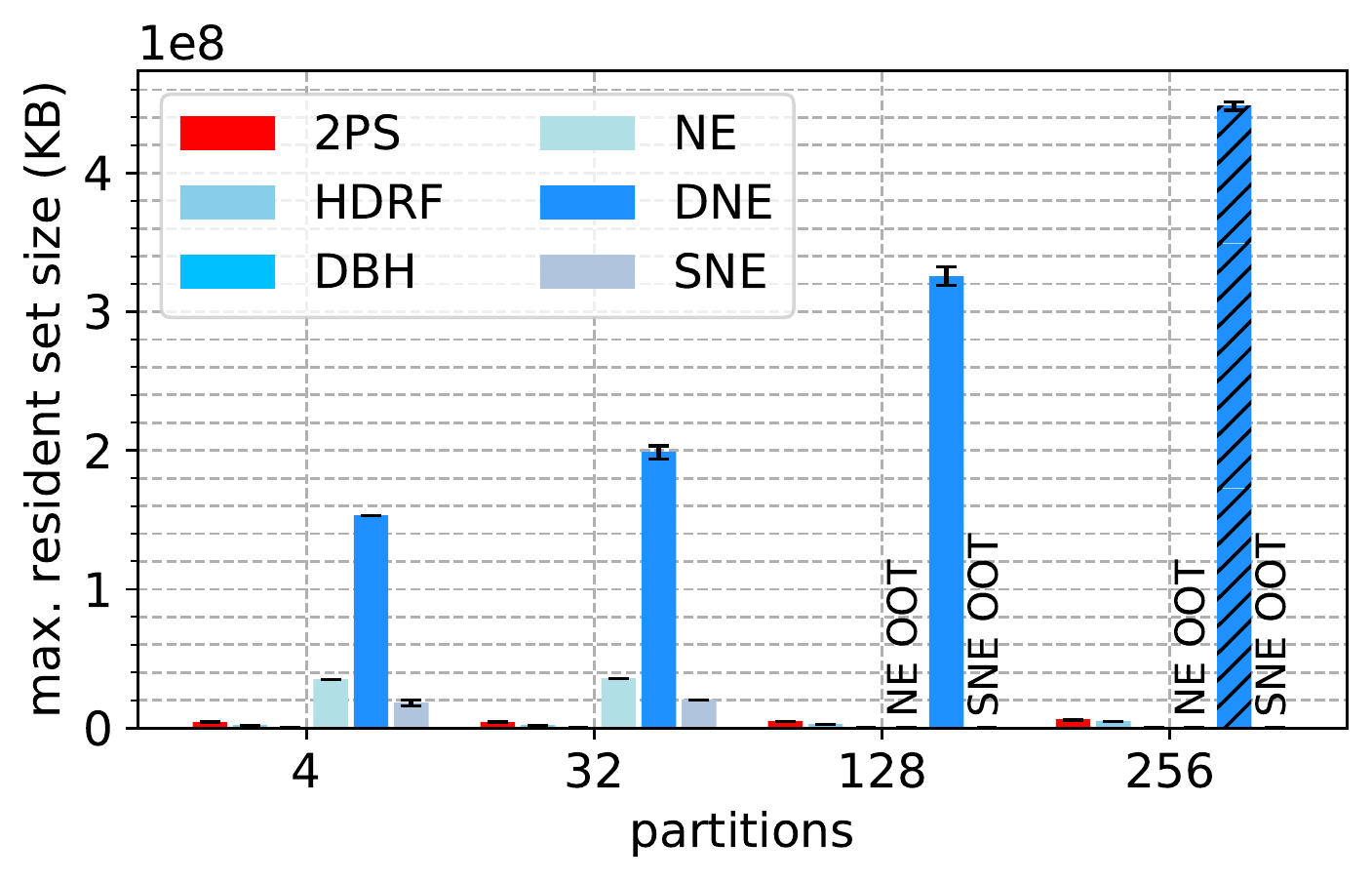}} \\
	\vspace{-0.45cm}
		\subfloat[UK: Replication factor.]{\label{a}   \includegraphics[width=0.295\textwidth]{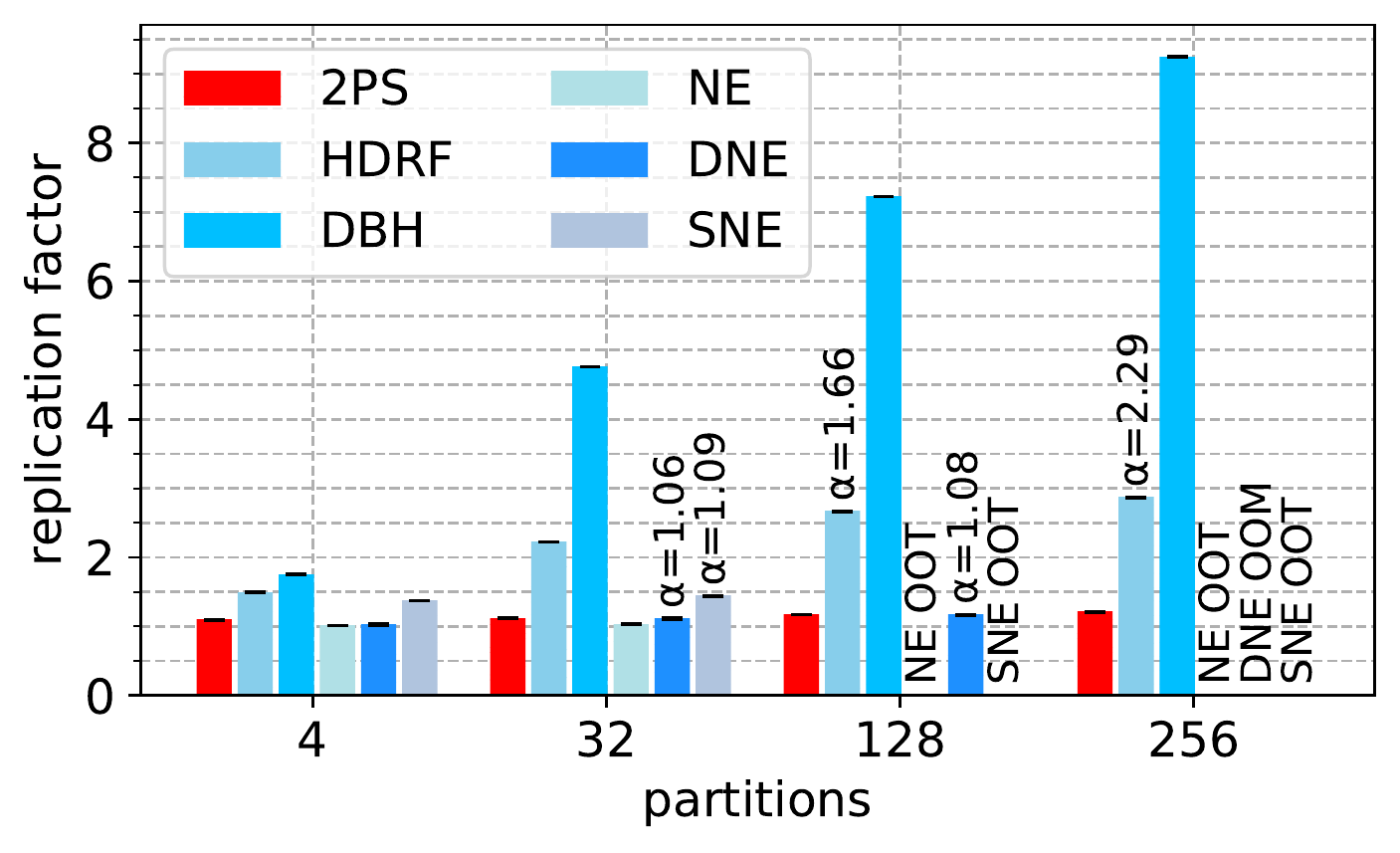}}
	\subfloat[UK: Run-time.]{\label{b}   \includegraphics[width=0.295\textwidth]{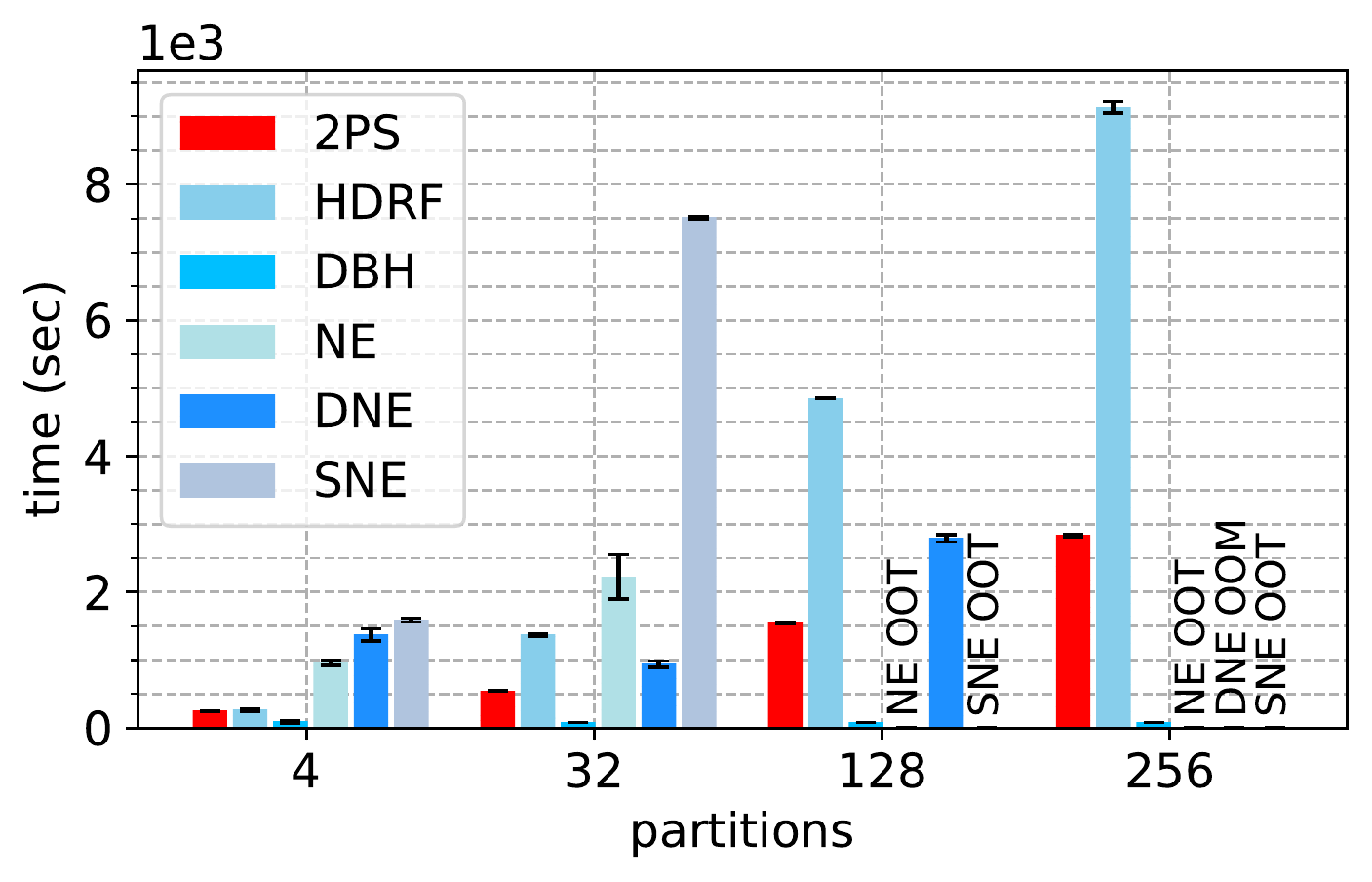}}
	\subfloat[UK: Memory overhead.]{\label{c}   \includegraphics[width=0.295\textwidth]{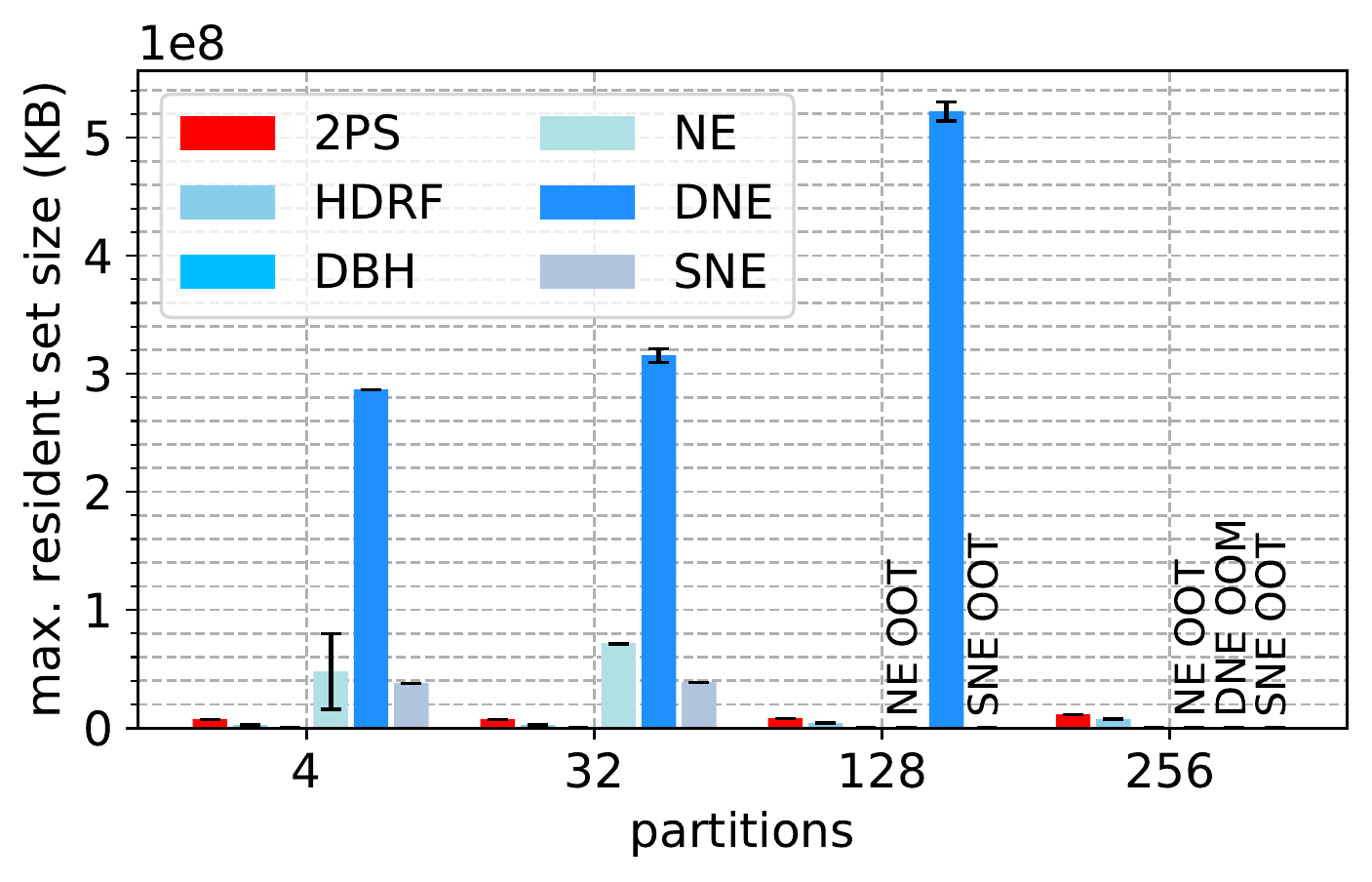}} \\
	\vspace{-0.45cm}
		\subfloat[GSH: Replication factor.]{\label{a}   \includegraphics[width=0.295\textwidth]{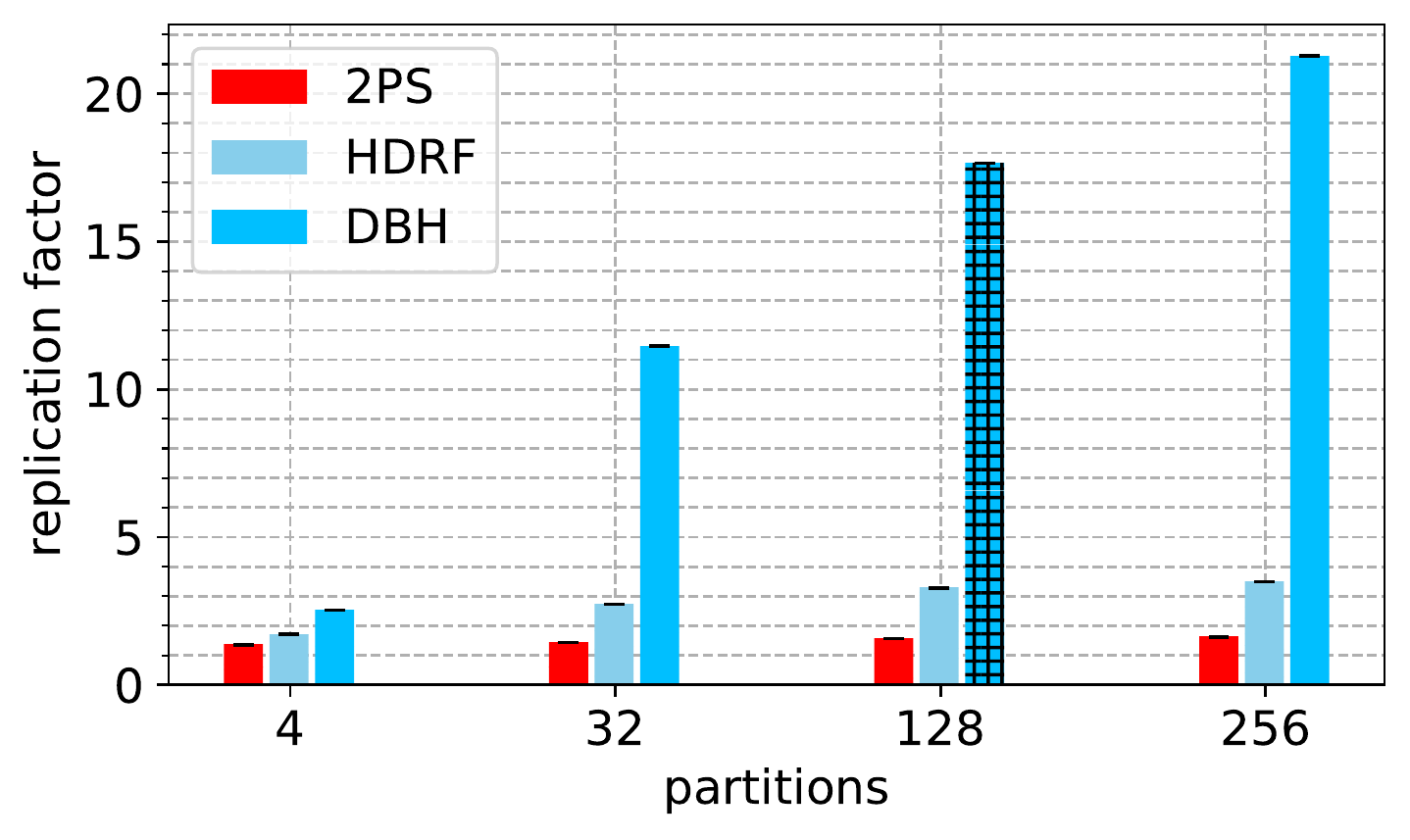}}
	\subfloat[GSH: Run-time.]{\label{b}   \includegraphics[width=0.295\textwidth]{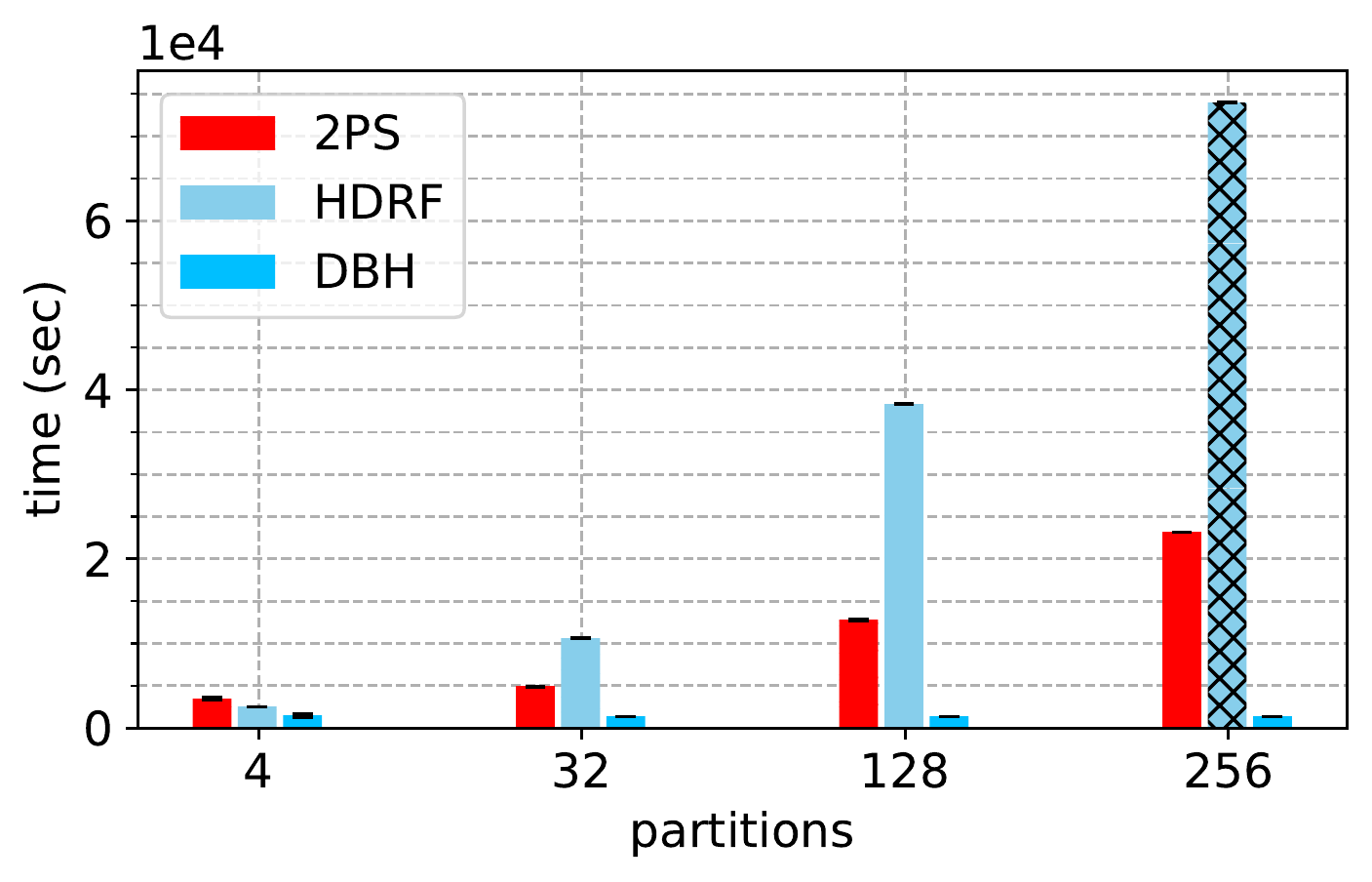}}
	\subfloat[GSH: Memory overhead.]{\label{c}   \includegraphics[width=0.295\textwidth]{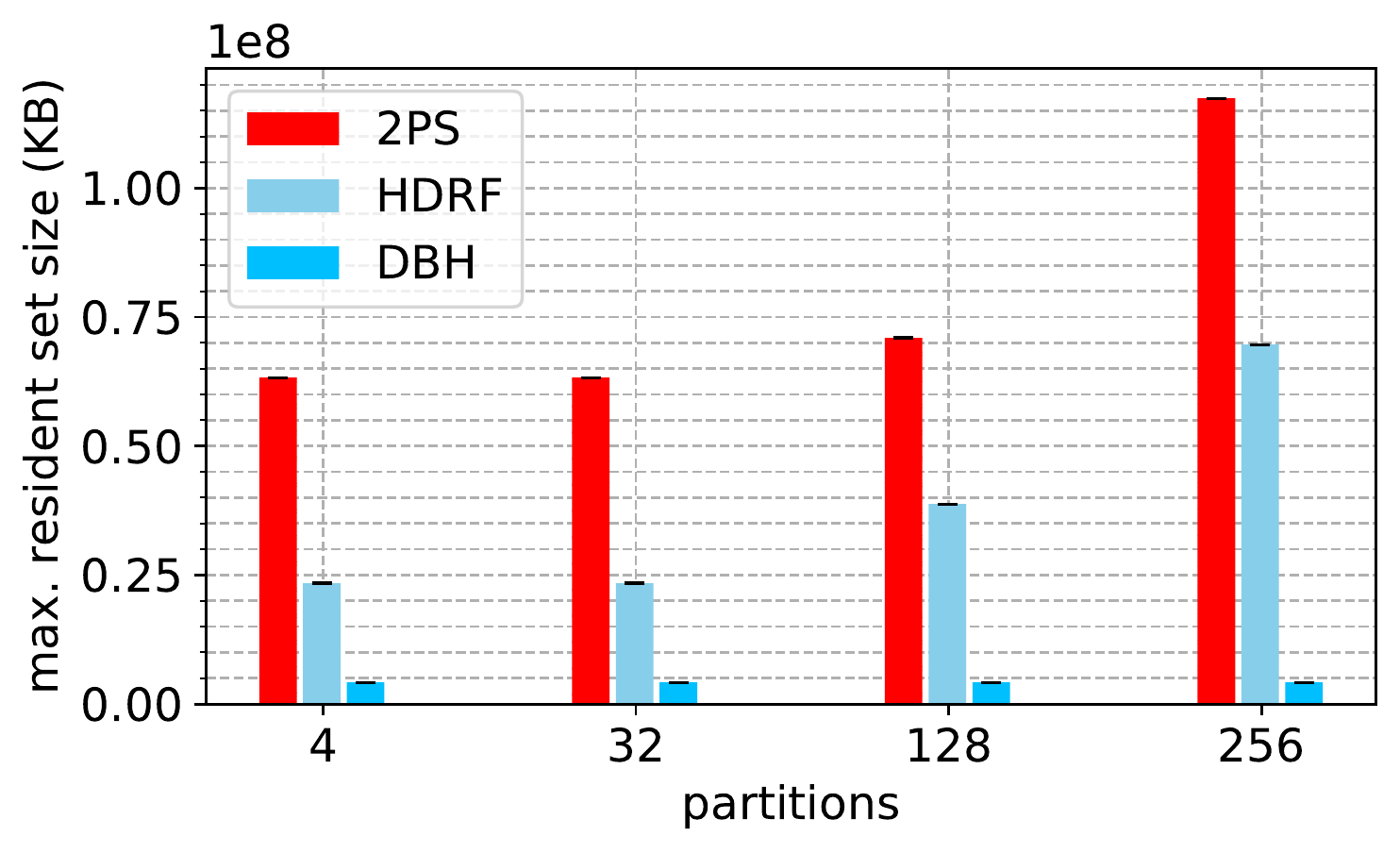}}\\
	\vspace{-0.4cm}
		\subfloat[WDC: Replication factor.]{\label{a}   \includegraphics[width=0.295\textwidth]{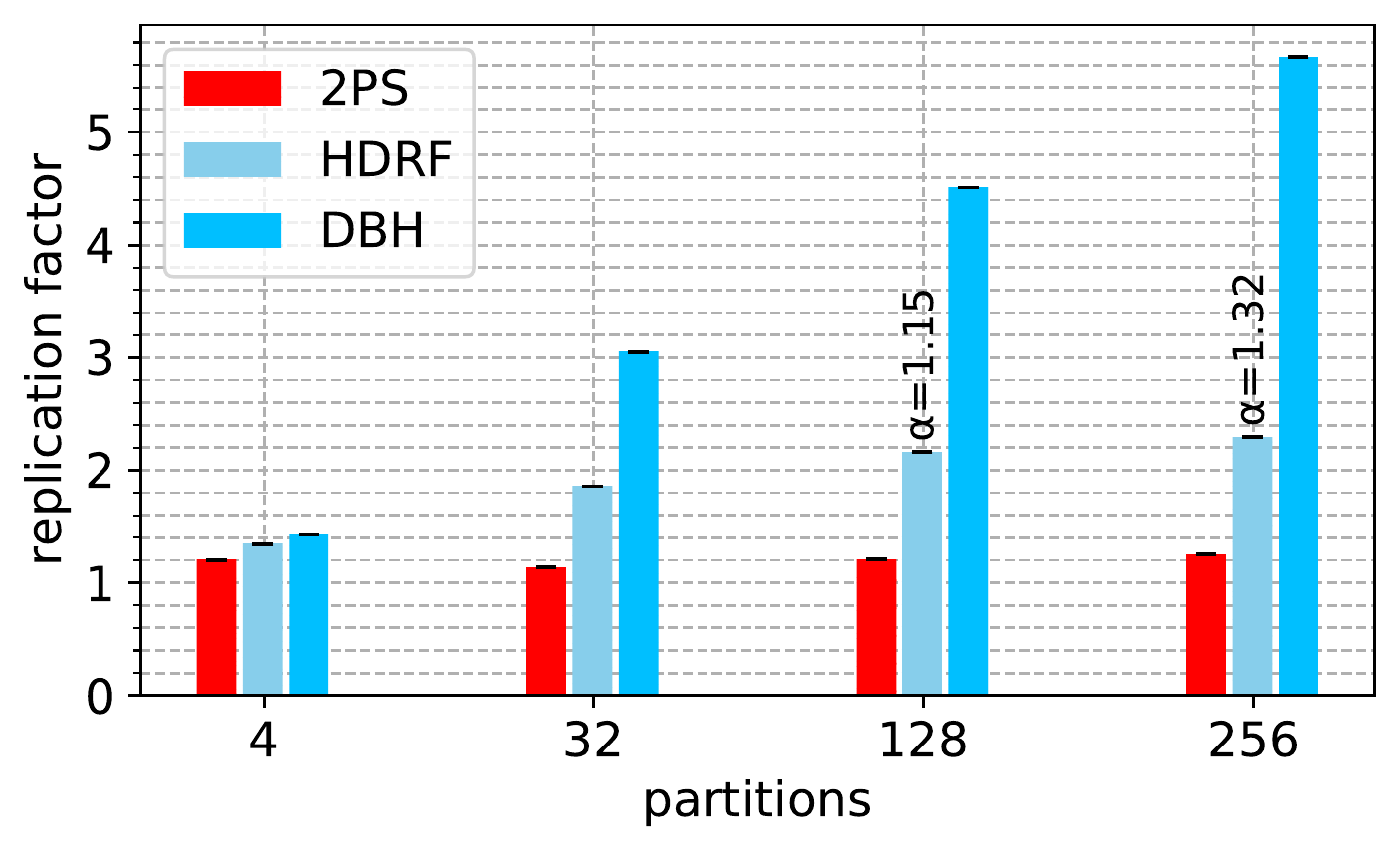}}
	\subfloat[WDC: Run-time.]{\label{b}   \includegraphics[width=0.295\textwidth]{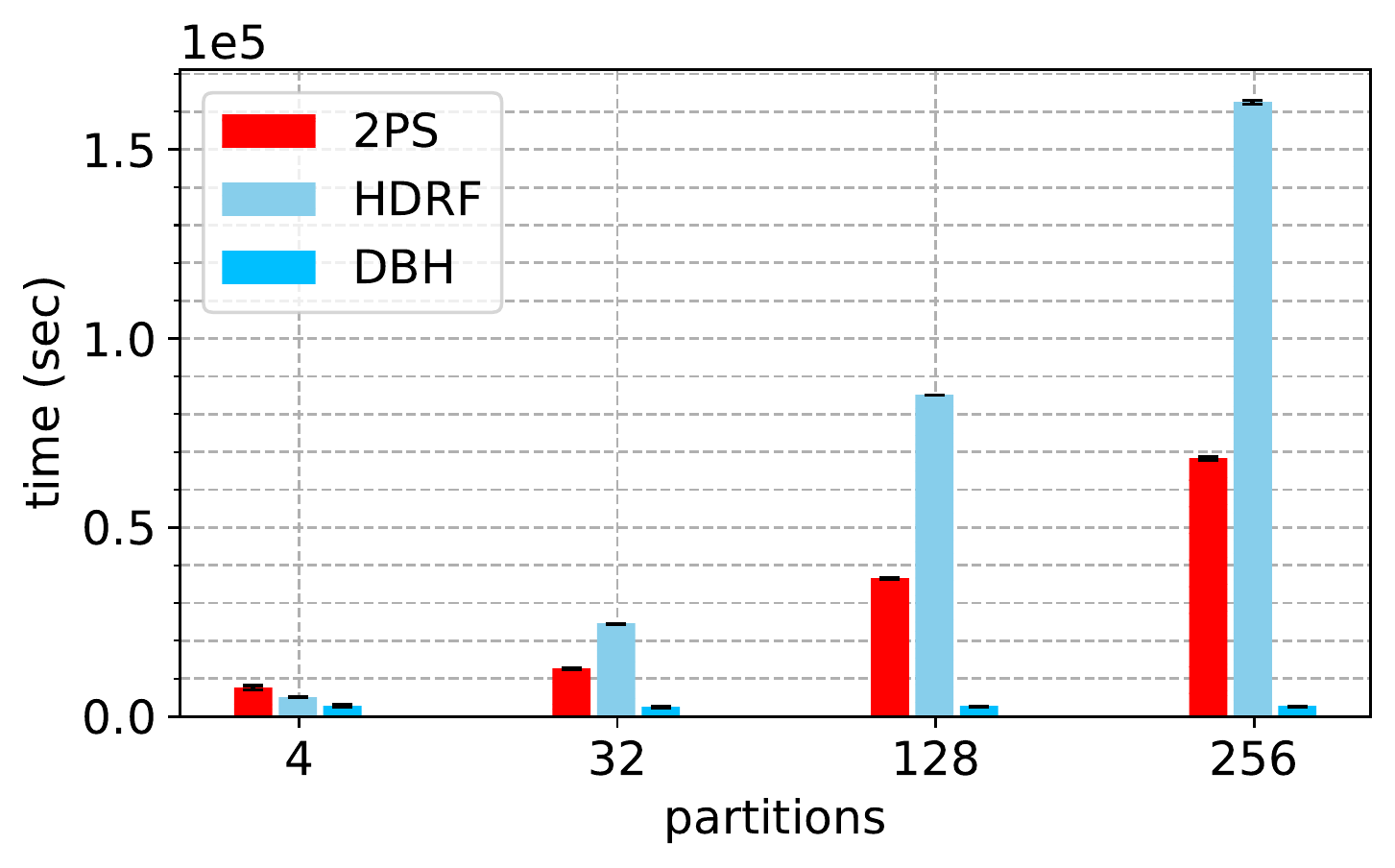}}
	\subfloat[WDC: Memory overhead.]{\label{c}   \includegraphics[width=0.295\textwidth]{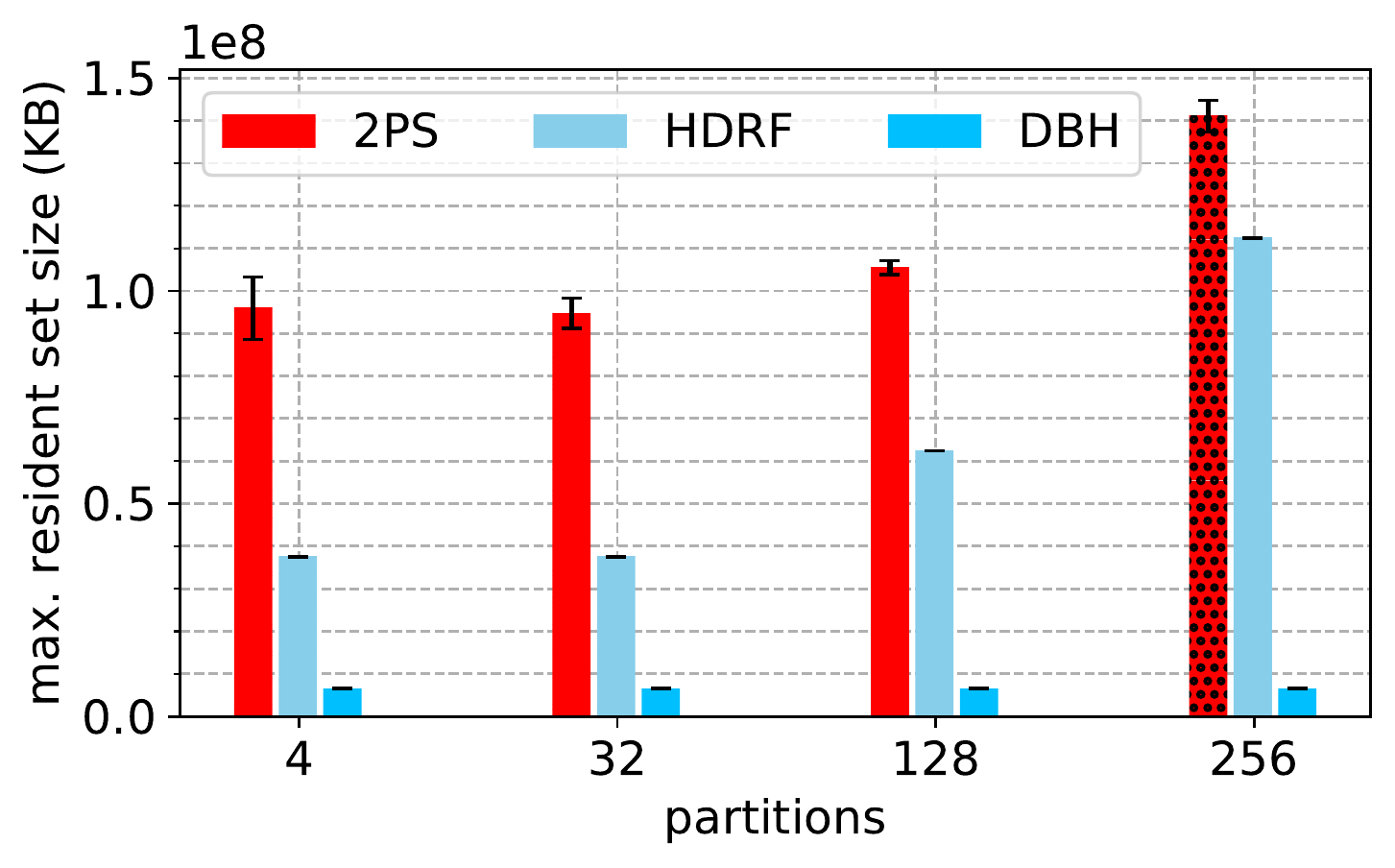}}
	\vspace{-5pt}
	\caption{Performance results on real-world graphs. OOT = out of time; OOM = out of memory.}
	\label{eval:perf}
%	\vspace{-16pt}
\end{figure*}

\paragraph{Main Observations}
In Figure~\ref{eval:perf}, we depict all performance measurements. Our main observations are as follows.

\textbf{(1) 2PS outperforms HDRF in terms of replication factor \emph{and} run-time.} Throughout all graphs and all numbers of partitions, 2PS yields a significantly lower replication factor than HDRF. This is in line with our theoretical result (see Section~\ref{sec:rf}). The advantage of 2PS is particularly pronounced on the web graphs (IT, UK, GSH and WDC). In terms of run-time, 2PS performs similar to HDRF on social network graphs (TW and FR) and is significantly faster on web graphs (IT, UK, GSH and WDC), even though 2PS performs multiple passes through the edge list, while HDRF only performs a single pass. This has two reasons. First, the clustering and pre-partitioning passes of 2PS are very lightweight. In particular, they do not involve computing a scoring function. Second, the more edges are pre-partitioned in 2PS, the fewer edges need to be partitioned via stateful streaming partitioning, which is more heavyweight due to the computation of a scoring function for every partition. In terms of memory overhead, the clustering state in 2PS requires a moderate amount of additional memory. However, this overhead is independent of the number of edges and the number of partitions. With a growing number of partitions, the relative memory overhead of 2PS compared to HDRF is diminishing. In comparison to NE and DNE, the overall memory overhead of 2PS is still negligible. 

\textbf{(2) 2PS outperforms SNE in terms of replication factor, run-time \emph{and} memory overhead.} Compared to 2PS, SNE induces a much larger run-time, which explodes when the number of partitions is high. For 128 or more partitions, we were not able to obtain results from SNE. Different from 2PS, SNE is not a real streaming algorithm, but instead implements a random-access partitioner (NE) with a partial view of the graph (i.e., it ingests and partitions subsequent chunks of the edge set). For every chunk of edges, an internal graph representation is built. This slows down the partitioning process. Only in one experiment (FR with 32 partitions), SNE could achieve a better replication factor than 2PS. However, in this specific case, the balancing constraint was severely violated, i.e., SNE built a heavily imbalanced partitioning. In the web graphs, despite of using more memory and inducing more run-time, the replication factor of SNE was always higher than the replication factor of 2PS. Finally, for GSH and WDC, we were not able to obtain results from SNE due to excessive run-time.

\begin{figure*}[]
	\centering
	
	%\subfloat[CAPTION]{BILDERCODE}\qquad
	\subfloat[Modularity after streaming clustering.]{\label{a}   \includegraphics[width=0.32\textwidth]{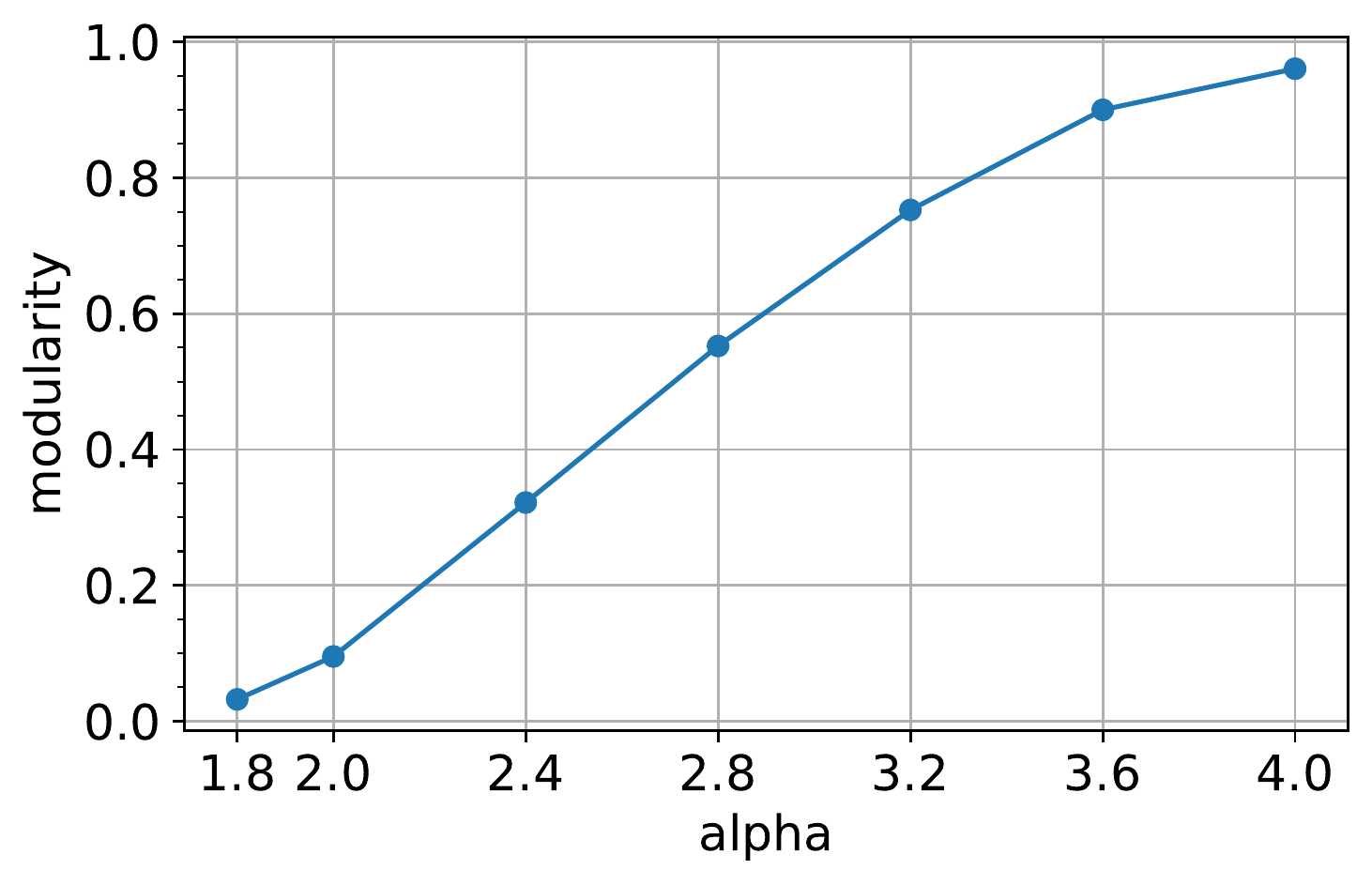}}
	\subfloat[Ratio of pre-partitioned edges.]{\label{b}   \includegraphics[width=0.32\textwidth]{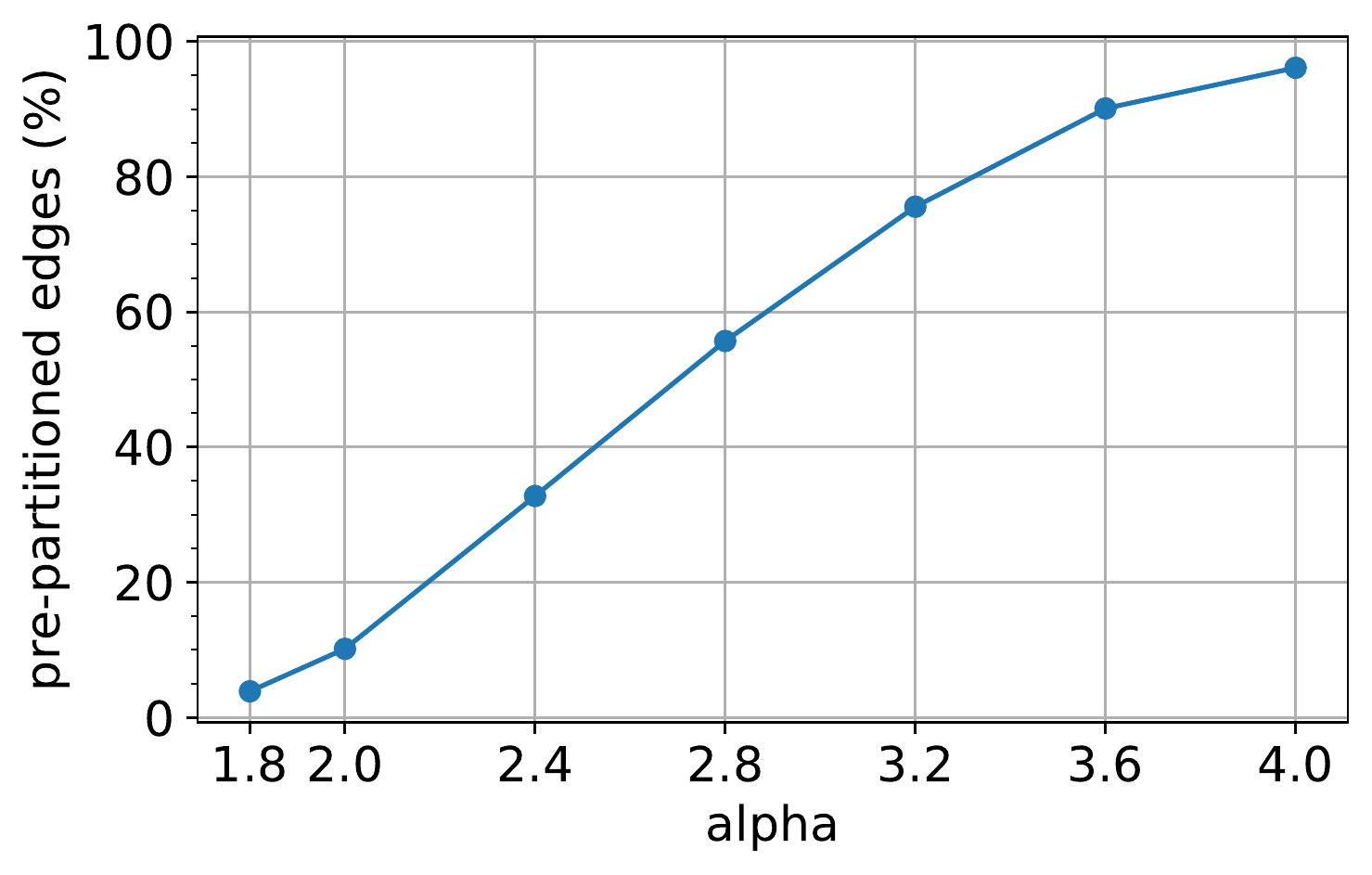}}
	\subfloat[Replication factor.]{\label{c}   \includegraphics[width=0.32\textwidth]{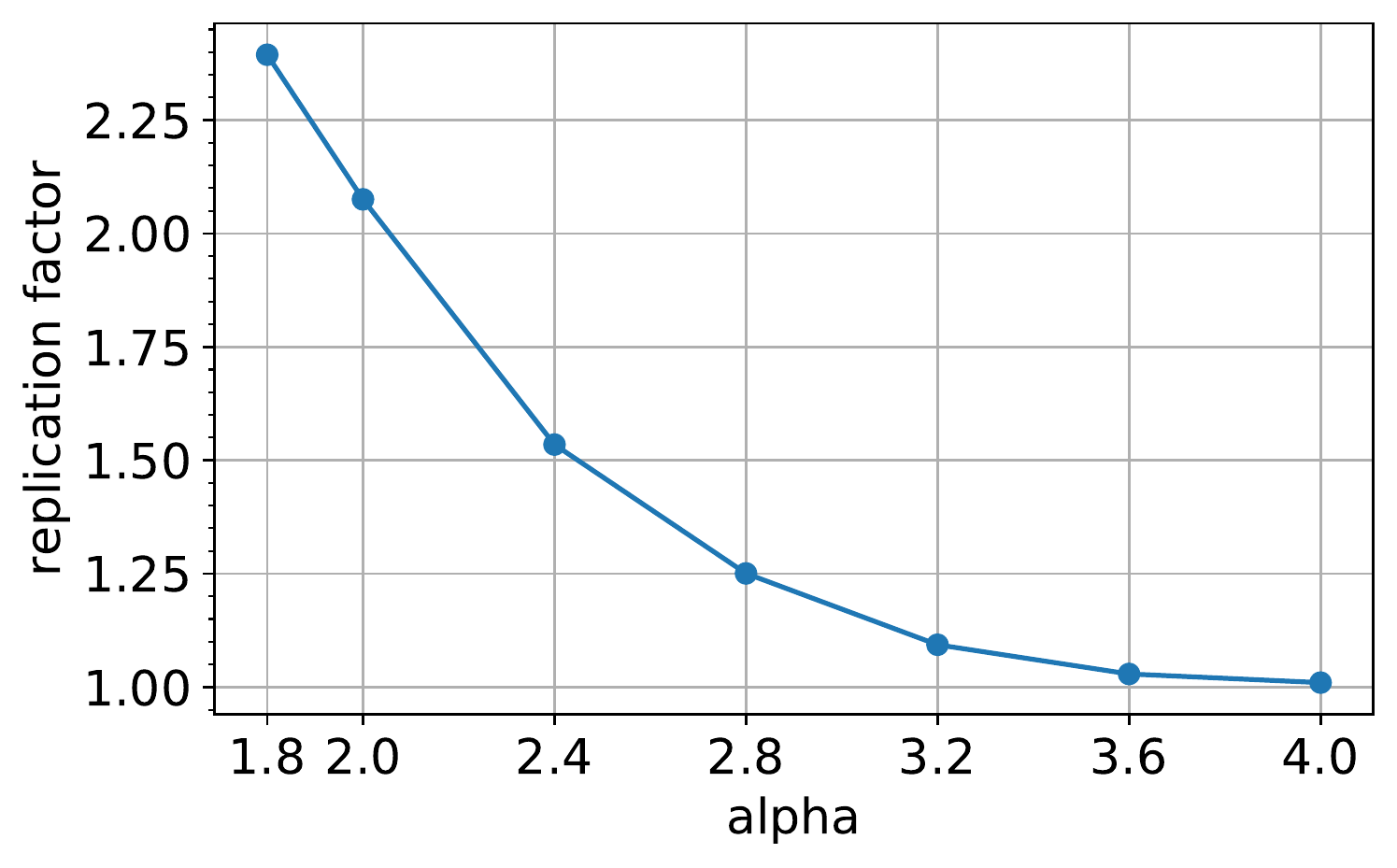}}
	\caption{Evaluations on synthetic graphs with different power-law degree exponent \texttt{alpha} for $k=128$ partitions.}
	\label{eval:synthetic}
%	\vspace{-16pt}
\end{figure*}

\textbf{(3) On web graphs, the replication factor of 2PS  is competitive to the random-access partitioners NE and DNE, but at a lower run-time and a lower memory overhead.} 2PS is capable of achieving a replication factor that is competitive to NE and DNE on web graphs (IT and UK), even though it is a streaming partitioner with an extremely low memory overhead. Moreover, 2PS is faster than NE on all graphs, and faster than DNE on the web graphs. Finally, due to its high memory efficiency, 2PS could successfully partition the large web graphs GSH and WDC, while NE and DNE ran out of memory.

\paragraph*{Further Discussion}
DBH produced the highest replication factor among all partitioners. On the other hand, DBH was also the fastest partitioner and induced the lowest memory overhead. Does this make DBH a good choice? Related studies show that investing some additional run-time into high-quality partitioning with low replication factor (compared to simple hashing, like in DBH) pays off when considering the total run-time of both graph partitioning and subsequent graph processing~\cite{Zhang:2017:GEP:3097983.3098033, 8416335}. Hence, using DBH is not the best option, especially when subsequent distributed graph processing is performed for many iterations. Different from DBH, which is fast but induces a \emph{high} replication factor, 2PS truly improves the state of the art in streaming edge partitioning by being \emph{faster} and inducing a \emph{lower} replication factor than its main competitor HDRF, as discussed above.

There were a couple of experiments that failed for some of the baseline partitioners. These failures either originate in excessive memory consumption (out of memory error) or excessive run-time (out of time error). On all graphs, when the number of partitions is large, NE and SNE show excessive run-time while DNE induces a large memory overhead. For the large web graphs GSH and WDC, NE and DNE both run out of memory, while SNE shows excessive run-time. In comparison to these baselines, the performance of 2PS is relatively stable. There is only a moderate increase of run-time and memory overhead with a growing number of partitions. Besides HDRF and DBH, 2PS is the only partitioner that could successfully partition all graphs with all numbers of partitions on our evaluation platform.

Finally, balancing was a surprisingly big challenge for some of the baseline partitioners. Especially for larger numbers of partitions, HDRF, NE, DNE and SNE had problems to keep the balancing constraint, even though NE, DNE and SNE explicitly offer a parameter to configure the maximum allowed imbalance. For HDRF, stricter balancing could potentially be achieved by setting a higher value of the parameter $\lambda$ which weighs in the balancing score in the scoring function. However, it is unclear what setting of $\lambda$ would work in which case. Different from these baselines, 2PS implements an explicit, hard volume cap for partitions. In other words, regardless of the shape and properties of the input graph, 2PS will always keep the configured balancing constraint.

In summary, 2PS is a major improvement of streaming edge partitioning. Different from other works, like ADWISE~\cite{8416335}, that just trade more run-time for a lower replication factor, we could reduce both replication factor \emph{and} run-time of streaming partitioning at the same time. Finally, by overcoming the uninformed assignment problem, 2PS proves to be a serious competitor in terms of replication factor even for random-access partitioners. At the same time, 2PS shows high efficiency, an aspect of data processing systems that is often overlooked~\cite{cost}.

\subsection{Impact of Power-Law Degree}

We further perform a set of experiments on synthetic graphs generated with the SNAP system\footnote{https://snap.stanford.edu/snap/download.html} (100.000 vertices, using the ``random-power law'' graph generator). In particular, we explore how the power-law degree exponent \texttt{alpha} of a random power-law graph\footnote{By convention, both the power-law degree exponent and the imbalance factor are often denoted $\alpha$ in the literature. To differentiate them without introducing divergent notation, we denote the imbalance factor  $\alpha$ and the power-law degree exponent of graphs \texttt{alpha}.} relates to the partitioning quality achieved with 2PS. We measure (a) the modularity of graph clustering achieved by the first phase of 2PS, (b) the ratio of pre-partitioned edges (i.e., edges assigned to partitions because of the clustering of their incident vertices), and (c) the overall replication factor. 

Figure~\ref{eval:synthetic} depicts the results. We make the following three observations:

(1) The higher the power-law degree exponent \texttt{alpha} of the graph, the better streaming clustering works. In a graph with $\mathtt{alpha} = 4.0$, the modularity achieved with streaming clustering is almost 1, which is the maximum possible modularity. This indicates that the graph is strongly clustered, and that the clustering algorithm has found these clusters.

(2) There is a direct relation between the modularity of streaming clustering and the ratio of pre-partitioned edges. In other words, the better the clustering works, the more edges can be assigned to partitions based on the clustering information, and the fewer edges have to be assigned to partitions based on stateful streaming partitioning. At $\mathtt{alpha} = 4.0$, almost 100~\% of the edges are assigned to partitions based on the clusters of their incident vertices. Therefore, the approach of applying modularity-based clustering in the first phase of 2PS is well justified.

(3) The higher the modularity of clustering and the ratio of pre-partitioned edges, the lower is the replication factor. At $\mathtt{alpha} = 4.0$, the achieved replication factor of 2PS is almost at 1.0, which is the best possible replication factor in edge partitioning.

Our study on synthetic graphs indicates that the power-law degree exponent is an important factor in the effectiveness of 2PS partitioning. According to a meta study by Aiello et al.~\cite{aiello2002random}, the observed power-law degree exponents of real-world web graphs have been reported to be between 2.1 and 2.7 in different studies. Hence, web graphs are expected to be in a range of power-law degree exponents where moderate modularity can be achieved by 2PS clustering. Indeed, our performance results (cf. Section~\ref{sec:realworld}) indicate a high effectiveness of 2PS on power-law graphs.

%-------------------------------------------------------------------------------
%-------------------------------------------------------------------------------
\section{Related Work}
\label{sec:related}

Graph partitioning is a problem with a long history in research~\cite{gp-survey}. It has numerous applications in solving optimization problems, e.g., in VLSI design~\cite{Karypis:1998:FHQ:305219.305248} and operator placement in stream processing systems~\cite{10.1007/978-3-642-10445-9_16}. In this paper, we focus on the edge partitioning problem. In particular, edge partitioning has gained a lot of attention as a preprocessing step for distributed graph processing systems, e.g., PowerGraph~\cite{powergraph} and Spark/GraphX~\cite{graphx}. 

Random-access partitioners~\cite{Zhang:2017:GEP:3097983.3098033, Margo:2015:SDG:2824032.2824046, dne}  require full information about the graph structure, i.e., the complete graph is loaded into memory of either a single machine~\cite{Zhang:2017:GEP:3097983.3098033} or a cluster of multiple machines~\cite{Margo:2015:SDG:2824032.2824046, dne} before partitioning is performed. The major shortcoming of such partitioners is that they consume a lot of memory. However, memory is a scarce and expensive resource. While a single machine may not have sufficient memory capacities to keep a large graph, employing a cluster of machines to perform graph partitioning induces a higher monetary cost. Streaming edge partitioning is a common approach to reduce the memory overhead of edge partitioning.

There are a number of approaches to streaming edge partitioning. Besides stateless partitioning based on hashing (e.g., DBH~\cite{dbh}), stateful partitioners have received growing attention. The basic idea is that by gathering state about past assignment decisions, future edges in the stream can be partitioned with a lower replication degree. In this regard, the main differentiation between streaming partitioners so far has been in the formulation of the scoring function. Different such functions have been proposed, e.g., Greedy~\cite{powergraph} and HDRF~\cite{Petroni:2015:HSP:2806416.2806424}. However, all of these approaches suffer from the uninformed assignment problem. An earlier approach to overcome this problem has been proposed in ADWISE~\cite{8416335}, which allows for dynamically re-ordering the edge stream such that the assignment of uninformed edges can be delayed. This way, locality in the edge stream can be exploited. However, the gains in replication factor depend on the window size, and a larger window imposes a larger run-time. Opposed to that, 2PS can reduce both replication factor and run-time compared to stateful streaming partitioning because the pre-partitioning based on vertex clustering is extremely lightweight and does not involve the computation of a scoring function. 

Stanton and Kliot~\cite{Stanton:2012:SGP:2339530.2339722} proposed a streaming model for vertex partitioning; the difference to streaming edge partitioning is that in streaming vertex partitioning, adjacency lists of vertices are ingested and vertices are assigned to partitions. FENNEL~\cite{Tsourakakis:2014:FSG:2556195.2556213} is a more recent partitioner that follows the vertex partitioning model. Bourse et al.~\cite{Bourse:2014:BGE:2623330.2623660} describe a method to transform vertex partitioning into edge partitioning. However, it requires significant preprocessing to create adjacency lists from an input edge list, as well as significant postprocessing to transform vertex partitioning into edge partitioning. These costs are avoided when the edge partitioning problem is solved directly, as we do in 2PS.

Some graph processing systems employ special execution models that require a specific style of graph partitioning different from edge partitioning or vertex partitioning. PowerLyra~\cite{powerlyra} is a hybrid distributed graph processing system that uses a combination of vertex-centric and edge-centric processing. Therefore, it also employs its own hybrid partitioning strategy that combines vertex partitioning and edge partitioning. Out-of-core graph processing systems like GraphChi~\cite{graphchi}, GridGraph~\cite{gridgraph} and Mosaic~\cite{Maass:2017:MPT:3064176.3064191} employ different graph preprocessing methods that involve specific graph partitioning problems.

A related problem is the partitioning of hypergraphs. In hypergraphs, an edge (sometimes called hyperedge) can connect more than only two vertices. This way, group relationships can be modeled. The problem of partitioning hypergraphs into multiple components while minimizing the cut size has also been tackled by streaming~\cite{Alistarh:2015:SMH:2969442.2969452} as well as random-access algorithms~\cite{1639359, TRIFUNOVIC2008563, Kabiljo:2017:SHP:3137628.3137650, 8621968}.

%-------------------------------------------------------------------------------%-------------------------------------------------------------------------------
\section{Conclusions}
\label{sec:conclusions}
In this paper, we tackle the uninformed assignment problem of stateful streaming edge partitioning and propose a new two-phase streaming approach. We make use of the great flexibility of graph clustering in the first phase, before we finalize and materialize the actual partitioning in the second phase. This way, we achieve state-of-the-art results in replication factor while staying faithful to a streaming model that does not require to keep the complete graph in memory at any time.
\bibliographystyle{plain}
\bibliography{\jobname}

%%%%%%%%%%%%%%%%%%%%%%%%%%%%%%%%%%%%%%%%%%%%%%%%%%%%%%%%%%%%%%%%%%%%%%%%%%%%%%%%
\end{document}